\documentclass[pre,twocolumn,showpacs,amsmath,amssymb,floatfix]{revtex4}
\usepackage{amsfonts} 
\usepackage{graphicx}
\usepackage{dcolumn}
\usepackage{bm}
\usepackage{comment}
\usepackage{epsfig}
\usepackage{times}
\usepackage[dvips]{color}  
\usepackage{color}
\usepackage{psfrag}
\usepackage{times}
\usepackage{graphics}
\usepackage{setspace}
\usepackage{latexsym}
\usepackage{amssymb}
\usepackage{amsmath}
\usepackage[tight]{subfigure}
\usepackage{comment}
\newcommand{\Label}[1]{\label{#1}}
\newcommand{\draftonly}[1]{}%
\newcommand{\DRAFT}%
{%
\renewcommand{\Label}[1]{\label{##1}{\hbox to 0cm{\textcolor{magenta}{\hss\em $##1$\quad}}}}}

\newcommand {\etal} {{\em et al.\/}}
\newcommand{\half}{\textstyle{\frac{1}{2}}}
\newcommand{\quarter}{\textstyle \frac{1}{4}}

\newcommand{\ashcomment}[1]{}

\def\d{d}
\def\v{{\bf v}}
\def\x{{\bf x}}
\def\eff{_\mathrm{eff}}

\def\th1{{\tilde\theta}}

\def\ncond{n_0}

\def\otheta{\hat{\mathrm{\theta}}}
\def\taux{\tilde{t}}
\def\waux{\tilde{\omega}}

\def\obk{\hat{b}_{\mathbf{k}}}
\def\obkdagg{\hat{b}^{\dagger}_{\mathbf{k}}}

\newcommand{\refb}[1]{(\ref{#1})}

\def\chik{\chi_{\mathbf{k}}}

\def\eff{_\mathrm{eff}}
\def\otheta{\hat{\mathrm{\theta}}}
\def\Dtwo{\widetilde{D}_{\mathsf{2}}}
\newcommand{\outss}{\mathrm{out}}
\newcommand{\inss}{\mathrm{in}}
\newcommand{\expanding}{\mathrm{exp}}
\def\taux{\tilde{t}}
\def\waux{\tilde{\omega}}

\begin{document}
%%%%%%%%%%%%%%%%%%%%%%%%%%%%%%%%%%%%%%%%%%%%%%%%%%%
\title{Analogue model of a FRW universe in Bose--Einstein condensates:\\
Application of the classical field method}%
\author{Piyush Jain}%
\email{piyushnz@gmail.com}
\affiliation{School of Chemical and Physical Sciences, Victoria University of Wellington, New Zealand}%
\author{Silke Weinfurtner}%
\email{silke.weinfurtner@mcs.vuw.ac.nz}
\author{Matt Visser}%
\email{matt.visser@mcs.vuw.ac.nz}
\affiliation{School of Mathematics, Statistics, and Computer Science, Victoria University of Wellington, New Zealand}%
\author{C.~W. Gardiner}
\email{gardiner@physics.otago.ac.nz}
\affiliation{The Jack Dodd and Dan Walls Centre for Photonics and Ultra Cold Atoms, Department of Physics, Otago University, Dunedin, New Zealand}%
%%%%%%%%%%%%%%%%%%%%%%%%%%%%%%%%%%%%%%%%%%%%%%%%%%%
\date{19 February 2007; \LaTeX-ed \today}% It is always \today, today,
%%%%%%%%%%%%%%%%%%%%%%%%%%%%%%%%%%%%%%%%%%%%%%%%%%%
\begin{abstract}
Analogue models of gravity have been motivated by the possibility of investigating phenomena not readily accessible in their cosmological counterparts. 
In this paper, we investigate the analogue of cosmological particle creation in a Friedmann--Robertson--Walker universe by numerically simulating a Bose--Einstein condensate with a time-dependent scattering length. In particular, we
focus on a two-dimensional homogeneous condensate using the classical field method via the truncated Wigner approximation. We show that for various forms of the scaling function the particle production is consistent with the underlying theory in the long wavelength limit. In this context, we further discuss the implications of modified dispersion relations that arise from the microscopic theory of a weakly interacting Bose gas.
\end{abstract}
%%%%%%%%%%%%%%%%%%%%%%%%%%%%%%%%%%%%%%%%%%%%%%%%%%%
\pacs{Valid PACS appear here}
%%%%%%%%%%%%%%%%%%%%%%%%%%%%%%%%%%%%%%%%%%%%%%%%%%%
%%%%%%%%%%%%%%%%%%%%%%%%%%%%%%%%%%%%%%%%%%%%%%%%%%%
\maketitle
%%%%%%%%%%%%%%%%%%%%%%%%%%%%%%%%%%%%%%%%%%%%%%%%%%%
%
%%%%%%%%%%%%%%%%%%%%%%%%%%%%%%%%%%%%%%%%%%%%%%%%%%%
%
\section{Introduction} \label{sect:expandingintro}
%
%%%%%%%%%%%%%%%%%%%%%%%%%%%%%%%%%%%%%%%%%%%%%%%%%%%
In the theory of quantum fields in classical backgrounds, some form of particle creation is typically expected when the metric is time-dependent; a commonly cited example of this is cosmological particle production in an expanding (or contracting) universe \cite{Birrell1982, Fulling1989}. It is possible to simulate the analogy of this process in a Bose--Einstein condensate (BEC) when either:  the external trapping frequency is time-dependent \cite{Weinfurtner2004,Fedichev2003,Fedichev2003a,Fedichev2004,Uhlmann2005}; or the scattering length (within the low momentum approximation of the two-body interaction potential) is time-dependent \cite{Barcelo2003,Duine2002,Fischer2004a}. These treatments are often based on the acoustic (\emph{i.e.}, hydrodynamic) approximation where it is assumed that all excitations of the quantum field propagate as phonons with the same speed of sound. Moreover back-reaction of the excitations on the background field, and higher-order interactions between excitations, are neglected in the linearized theory.

In general, the correct description of the dynamics of a BEC is a formidable problem due to the vastness of the Hilbert space, even for a system of just a few interacting atoms.  In the lowest-order approximation, when all the bosonic atoms occupy a single quantum state, the ground state is well described by the Gross--Pitaevskii equation (GPE) --- in this case, the field operator is replaced by a mean-field order parameter. Classical field methods (CFM) extend this formalism to include quantum fluctuations whereby the dynamics of a multimode quantum field is approximated by the trajectories of classical variables in phase space. One such method is the truncated Wigner approximation (TWA), which is based on the Wigner representation of the density matrix. The TWA has been investigated by a number of authors \cite{Steel1998,Sinatra2000,Sinatra2001,Sinatra2002} and more recently has been applied to a study of condensate collisions \cite{norrie2005,norrie2006}. 

In this paper we investigate the dynamics of a homogeneous BEC in two spatial dimensions with a time-dependent scattering length, and in particular, map this problem to a Friedmann--Robertson--Walker (FRW) universe undergoing an expansion. We compare the analytically calculated particle production from the acoustic approximation with the results of numerical simulations based on the TWA. There are several benefits to this approach. Firstly, the TWA includes the effects of vacuum quantum fluctuations by sampling the Wigner distribution in the initial state. Secondly, the field dynamics naturally include the nonlinear dispersion of the Bose system --- in a cosmological context, this represents Lorentz symmetry breaking of the effective spacetime, which leads to necessary modifications of the standard hydrodynamic theory. Finally, the numerical simulations include the effects of back-reaction, which is difficult to otherwise include without resorting to higher-order methods such as the self-consistent Hartree--Fock--Bogoliubov approach.

The outline of this paper is as follows:
In Sec.~\ref{sect:emergentFRW} we show how the acoustic metric leads to an emergent FRW universe in a BEC, which leads to the prediction of quasiparticle production as discussed in Sec.~\ref{sect:quantisation}. In Sec.~\ref{sect:bogconnection} the these ideas are formalised by the Bogoliubov theory for a BEC. In Secs.~\ref{sect:acousticpredictions} and \ref{sect:nonlindispresults} respectively the preceding theory is used to quantitatively predict quasiparticle production in the acoustic approximation, and in the more general case including ``trans-phononic'' modes, for a number of specific FRW universe models ~\cite{Weinfurtner:2007ab}.
Section \ref{chap:cfm} introduces the TWA, which we use to simulate the dynamics of the BEC consistently with these scenarios. Section \ref{chap:expandingsimulations} presents the results of these simulations, which are discussed in Sec.~\ref{sect:discussion}. Finally, in Sec.~\ref{conclusions} we conclude and discuss avenues for further work.

%%%%%%%%%%%%%%%%%%%%%%%%%%%%%%%%%%%%%%%%%%%%%%%%%%%
%
\section{Expanding Universe Models in Bose--Einstein condensates}\label{sect:emergentFRW}
%
%%%%%%%%%%%%%%%%%%%%%%%%%%%%%%%%%%%%%%%%%%%%%%%%%%%
Arguably, the most promising system for implementing analogue models of gravity is the BEC. This possibility was first considered by Garay \etal~\cite{Garay2000,Garay2001} for acoustic black hole geometries, and further explored by Barcel\'{o}, Visser, and Liberati \cite{Barcelo2001,Visser2002,Barcelo2003b}. In particular, BECs have a number of desirable features that are necessary to be part of the analogue model programme, and more specifically for the subgroup of analogue models suitable for mimicking cosmological particle production:
\begin{enumerate}
\item[(i)] Hydrodynamics: In the long-wavelength limit, the mean-field equations of motion for a BEC take the form of classical hydrodynamics for a superfluid, which leads directly to the formulation of an emergent spacetime. 
\item[(ii)] Quantum theory: Bose--Einstein condensation in atomic vapours is a weakly interacting system, for which the microscopic quantum theory is well understood. The elementary excitations of the system are given by the Bogoliubov theory. The description for the excitations is valid beyond the hydrodynamic approximation, and incorporates ``trans-phononic'' physics, similar to many (not all) effective field theories, where ``trans-Planckian'' physics is believed to break Lorentz invariance.
\item[(iii)]  Temperature: BECs in atomic vapours require temperatures close to absolute zero so that in principle it may be possible to observe cosmological particle production, without the presence of (larger) thermal fluctuations that would obscure the effect.
\item[(iv)] Experimental advances: Recent experimental advances for the control of ultra-cold atoms mean that BECs can now be prepared and manipulated in many configurations. Notably, the use of magnetic and optical traps can lead to a variety of geometries, whereas using Feshbach resonances it is possible to vary the interaction strength between atoms, even by many orders of magnitude. With continued advances the experimental realization of analogue models should soon be achievable. 
\end{enumerate}

%+++++++++++++++++++++++++++++++++++++++++++++++++++++++++++++++++
\subsection{Microscopic theory of the Bose gas}\label{subsect:microscopic}
%+++++++++++++++++++++++++++++++++++++++++++++++++++++++++++++++++
Bose--Einstein condensation is characterised by a macroscopic occupation of a single quantum state, typically the ground state of the system. This quantum degeneracy is achieved at very low temperatures where the de Broglie wavelength becomes comparable to the inter-particle spacing.
In the second-quantised formalism, the effective Hamiltonian for a dilute Bose gas is
\begin{eqnarray}\label{secondquantH}
\hat{H} = \hat{H_0} + \hat{H_I} ,
\end{eqnarray}
where the single particle Hamiltonian is
\begin{eqnarray}
\hat{H_0} = \int d \x \, \hat{\psi^{\dagger}}(\x) \left [-\frac{\hbar^2}{2 m} \nabla^2 +
V_{\textrm{ext}}(\x) \right ] \hat{\psi}(\x),
\end{eqnarray}
and the interaction Hamiltonian is
\begin{eqnarray}
\hat{H_I} = \frac{U}{2} \int d\x \, \hat{\psi^{\dagger}}(\x) \hat{\psi^{\dagger}}(\x) \hat{\psi}(\x) \hat{\psi}(\x).
\end{eqnarray}
$V_{\scriptsize{\textrm{ext}}}(\x)$ is any external potential (\emph{e.g.}, from a trap) and the two-body potential has been approximated via a contact potential $U = 4\pi\hbar^2 a/m$ in terms of the $s$-wave scattering length $a$, valid in the cold collision regime. 
The field operator $\hat{\psi}(\x)$ annihilates a boson at position $\x$ and obeys the usual equal time commutation relations
\begin{eqnarray}\label{psicommutation}
[\hat{\psi}(\x, t), \, \hat{\psi}(\x^{\prime}, t)] 
&=& [\hat{\psi^{\dagger}}(\x, t), \, \hat{\psi^{\dagger}}(\x^{\prime}, t)] = 0, \\ \nonumber
[\hat{\psi}(\x, t), \, \hat{\psi^{\dagger}}(\x^{\prime}, t)] &=& \delta(\x - \x^{\prime}).
\end{eqnarray}
The corresponding Heisenberg equation of motion for the field operator is
\begin{eqnarray}\label{heisenbergfield}
i \hbar \frac{\partial \hat{\psi}(\x, t)}{\partial t} &=& [\hat{\psi}, \hat{H}] \nonumber \\ &=& \left [-\frac{\hbar^2}{2 m} \nabla^2 + V_{\scriptsize{\textrm{ext}}}  + U \hat{\psi^{\dagger}} \hat{\psi} \right ] \hat{\psi}.
\end{eqnarray}
Without further approximations, this equation cannot be solved for a realistic system since the Hilbert space becomes prohibitively large, even for a system of just a few atoms. 

For the case of a weakly interacting Bose gas, and for low temperatures $T \ll T_{\mathrm{critical}}$, a very useful approximation arises from the fact almost all the atoms reside in a single quantum state, so the system occupies only a fraction of the available quantum states. In this case it is useful to write
\begin{eqnarray}\label{eq:meanfield}
\hat{\psi}(\x, t) = \psi(\x, t) + \delta \hat{\varphi}(\x, t),
\end{eqnarray}
where $\psi(\x, t) = \langle \hat{\psi}(\x, t) \rangle$ is a mean-field term (the \emph{condensate wave function}) and $\delta \hat{\varphi}(\x, t)$ is that part of the quantum field associated with quantum and thermal fluctuations, with $\langle \delta \hat{\varphi}(\x, t) \rangle = 0$. This is known as the \emph{Bogoliubov approximation}.
This description itself may be treated with varying levels of approximation. In the very simplest approximation fluctuations are neglected altogether, and the resulting equation, known as the \emph{Gross--Pitaevskii equation} (GPE), is given by
\begin{eqnarray}\label{eq:gpe}
i \hbar \frac{\partial \psi(\x, t)}{\partial t}\!=\!\left[-\frac{\hbar^2}{2 m} \nabla^2\!+\! V_{\scriptsize{\textrm{ext}}}(\x)\!+\!U |\psi(\x, t)|^2 \right ] \psi(\x, t).
\end{eqnarray}
This equation provides a description of the condensate in terms of a classical field.

%+++++++++++++++++++++++++++++++++++++++++++++++++++++++++++++++++
\subsection{Establishing the analogy}\label{subsect:analogy}
%+++++++++++++++++++++++++++++++++++++++++++++++++++++++++++++++++
It is now well established that effective spacetimes emerge from the microscopic theory of BECs in the long wavelength limit \cite{Barcelo2001, Barcelo2006}. 
In particular, the analogy between curved spacetime and BECs can be revealed starting from the mean-field description of condensates, given by the GPE (\ref{eq:gpe}). The complex order parameter $\psi(\mathbf{x}, t)$ can be written in terms of a real density and phase (\emph{i.e.}, the \emph{Madelung} representation) as
\begin{eqnarray}
\psi(\mathbf{x}, t) = \sqrt{n(\mathbf{x}, t)} e^{i \theta(\mathbf{x}, t)}.
\end{eqnarray}
Substituting this into (\ref{eq:gpe}) and equating real and imaginary parts leads to the pair of equations
\begin{eqnarray}\label{density}
\frac{\partial n}{\partial t} + \frac{\hbar}{m} \nabla(n \nabla \theta) = 0,
\end{eqnarray}
and
\begin{eqnarray}\label{phase}
\hbar \frac{\partial \theta}{\partial t} - \frac{\hbar^2}{2 m} \frac{\nabla \sqrt{n}}{\sqrt{n}} 
+ V_{\textrm{ext}} + U n + \frac{\hbar^2}{2 m}(\nabla \theta)^2 = 0.
\end{eqnarray}
Linearizing about the fields $n$ and $\theta$ by setting
\begin{eqnarray}
n \rightarrow n_0 + \hat n_1, \nonumber \\
\theta \rightarrow \theta_0 + \hat\theta_1,
\end{eqnarray}
equations (\ref{density}) and (\ref{phase}) then lead to
\begin{eqnarray}\label{n1_linearized}
\frac{\partial \hat n_1}{\partial t} = \frac{\hbar}{m} (\hat n_1 \nabla \theta_0 + n_0 \nabla \hat \theta_1),
\end{eqnarray}
\begin{eqnarray}\label{theta1_linearized}
\hbar \frac{\partial \hat \theta_1}{\partial t} + {\widetilde U} \hat n_1 + \frac{\hbar^2}{m} \nabla \theta_0 \nabla \hat \theta_1 = 0,
\end{eqnarray}
where we have only retained terms up to first order in $\hat n_1$ and $\hat \theta_1$, and where we have defined the operator ${\widetilde U}$ by  
\begin{eqnarray}
{\widetilde U}\hat  n_1 \equiv \left[ U - \frac{\hbar^2}{2 m} \Dtwo \right] \hat n_1.
\end{eqnarray}
The differential operator $\Dtwo$ has been introduced, which to first order, takes the form:
\begin{eqnarray}\label{D_2}
\Dtwo \hat n_1 = -\frac{1}{2}n_0^{-3/2} (\nabla^2 \sqrt{n_0}) \hat n_1 + \frac{1}{2 \sqrt{n_0}} \nabla^2 \left( \frac{\hat n_1}{\sqrt{n_0}} \right).
\end{eqnarray}
The operator $\Dtwo$ gives the first order correction from the inclusion of the quantum pressure term. 
Identifying the background velocity for the irrotational field as 
\begin{equation}
\mathbf{v} = \hbar/m \nabla \theta_{0}, 
\end{equation}
and combining (\ref{n1_linearized}) and (\ref{theta1_linearized}) yields a second-order differential equation for the perturbed phase, which can be written in the compact form \cite{Barcelo2001,Barcelo2006}:
\begin{eqnarray}\label{fieldequation_f}
\partial_{\mu} ( f^{\mu \nu} \; \partial_{\nu} \hat \theta_1 ) = 0,
\end{eqnarray}
where we have introduced the symmetric matrix
\begin{eqnarray}
f^{\mu \nu}(\mathbf{x}, t) = 
 \left[
\begin{array}{ccc}
f^{00} & \vdots & f^{0j}  \\
\ldots & & \ldots  \\
f^{i0} & \vdots & f^{ij}  \\
\end{array}
\right],
\end{eqnarray}
with components \footnote{There is an assumption here, that the linear operator ${\widetilde U}$ is invertible, which is true only if the null space of ${\widetilde U}$ is the zero vector space. This is certainly true for the homogeneous BEC as we will show in Sec.~\ref{sect:bogconnection}.}
{\allowdisplaybreaks
\begin{eqnarray}
f^{00} & = & -{\widetilde U}^{-1}, \nonumber \\[1ex]
f^{0j} & = & -{\widetilde U}^{-1} v^j, \nonumber \\[1ex]
f^{i0} & = & - v^i {\widetilde U}^{-1}, \nonumber \\[1ex]
f^{ij} & = & \frac{n_0}{m} \delta^{ij} - v^i {\widetilde U}^{-1} v^j.
\end{eqnarray}
}
Note we have used the standard nomenclature where the Greek indices run from $0$ to $d$  and the Roman indices from $1$ to $d$ for $d$ spatial dimensions.

In the acoustic (or hydrodynamic) approximation the quantum pressure term is neglected. In this case $\Dtwo$ is set to zero (\ref{fieldequation_f}) can be written more simply as  
\begin{eqnarray}\label{fieldequation_g}
\frac{1}{\sqrt{-g}} \; \partial_{\mu} \left( \sqrt{-g} \; g^{\mu \nu} \; \partial_{\nu} \hat \theta_1 \right) = 0
\end{eqnarray}
where $g_{\mu \nu}$ is the effective covariant metric tensor (with determinant $g$) given by
\begin{eqnarray}\label{ghydro}
g_{\mu \nu}(\mathbf{x}, t) = \left(\frac{n_0}{c}\right)^{\frac{2}{d-1}}
\left[
\begin{array}{ccc}
-(c^2 - v^2) & \vdots & -v_j  \\
\ldots & & \ldots  \\
-v_i & \vdots & \delta_{ij}  \\
\end{array}
\right].
\end{eqnarray}
We have used the fact that the speed of sound in a condensate is given by 
\begin{equation}
c^2 = U n_0/m.
\end{equation}
The equation for the phase fluctuations (\ref{fieldequation_g}) is formally analogous to the dynamics of a massless and minimally coupled scalar field in a curved space-time \cite{Birrell1982}.
 It is worth emphasising that although the external potential $V_{\textrm{ext}}$ does not explicitly appear here, the field equation still depends on $V_{\textrm{ext}}$ implicitly, as the the background geometry is determined by the stationary solutions of the GPE (\ref{eq:gpe}).

In the acoustic approximation all collective excitations behave as sound waves with the usual linear ``relativistic'' dispersion relation 
\begin{equation}
\omega = c \, k ; 
\end{equation}
the quanta of excitations are thus phonons. An interesting consequence of Bogoliubov theory in Bose--Einstein condensates is that in general the excitation spectrum displays nonlinear dispersion (see Sec.~\ref{sect:bogconnection}), being linear (\emph{i.e.}, phononic) for low $|\mathbf{k}|$ and becoming quadratic (\emph{i.e.}, free-particle like) at large $|\mathbf{k}|$. When nonlinear dispersion is incorporated into analogue models of gravity it is equivalent to breaking Lorentz invariance \cite{Barcelo2006,Visser:2001aa,Weinfurtner:2006iv}. We will return to this point in Sec.~\ref{lorentzbreak}, and discuss the implications of this for the analogue model in a companion paper \cite{Weinfurtner:2007ab}.

Equation (\ref{fieldequation_g}) is the massless Klein--Gordon equation. 
The metric (\ref{ghydro}) has the signature ($-$$+$$+$$+$) and so the effective spacetime represents a Lorentzian geometry. It is interesting to note that an effective ``relativistic'' wave equation (for excitations in the condensate) arises from the equations of motion for a non-relativistic fluid~\cite{Unruh:1981bi,Visser:1993tk}.

The hydrodynamic (\emph{i.e.}, long wavelength) description of a BEC is commonly believed to lead to an effective field theory (EFT) for curved spacetime in the same sense that classical gravity is expected to emerge as the low-energy EFT of a full theory of quantum gravity \cite{Donoghue1994}. (The analogue gravity EFT arising from a BEC does not impose a strict cut-off and so includes so-called \emph{trans-phononic} physics~\cite{Liberati:2005id,Weinfurtner:2006nl,Weinfurtner:2006iv,Weinfurtner:2006eq,Liberati:2006sj} --- we return to this point in Sec.~\ref{lorentzbreak}). However, it should be noted that the emergent spacetime is itself embedded in an low-energy EFT for the Bose system; here the cut-off is determined by the validity of replacing the two-body interaction potential by an effective potential that is characterised by the $s$-wave scattering length. In practice, the cut-off is enforced by the the use of a projector in mode space.

We further emphasise that the above analogy only holds for massless spin-0 particles; in general, it is possible to modify the formalism to include massive modes at the expense of dealing with a more complex configuration, \emph{e.g.}, a two-component BEC \cite{Visser:2004zn,Visser:2005ai,Liberati:2006sj,Weinfurtner:2006eq,Weinfurtner:2006nl,Weinfurtner:2006iv,Liberati:2005id}. Moreover, it is possible to develop an analogue model of gravity for spin-1 fields, \emph{e.g.}, in dielectric media \cite{Schutzhold2002}.

%+++++++++++++++++++++++++++++++++++++++++++++++++++++++++++++++++
\subsection{FRW analogue model}
%+++++++++++++++++++++++++++++++++++++++++++++++++++++++++++++++++
Time dependence can enter the metric tensor (\ref{ghydro}) in any of the parameters $\ncond$, $c$, and $\v$. We focus on the case where the background flow is zero ($\v = 0$), and the system is homogeneous with a density $\ncond$ that is constant through space. With this choice of parameters the effective metric (\ref{ghydro}) becomes
\begin{eqnarray}\label{frw_metric}
g_{\mu \nu} = \left( \frac{n_0}{c} \right)^{\frac{2}{d-1}} 
\left[
\begin{array}{ccc}
-c^2 & \vdots & 0  \\
\ldots & & \ldots  \\
0 & \vdots & \delta_{ij}  \\
\end{array}
\right].
\end{eqnarray}
This represents a Friedmann--Robertson--Walker spacetime \cite{Barcelo2003, Duine2002}. (Technically, a $k=0$ spatially flat FRW spacetime.) Such spacetime geometries are conformally flat, and at any particular time the spatial geometry is simply that of Euclidean $3$-space. The time dependence is contained entirely in the speed of sound given by
\begin{equation} \label{soundspeed}
c(t)^2 = \frac{U(t) n_{0}}{m} = \frac{4\pi \hbar^2}{m^2} \, \,  n_{0} \, a(t),
\end{equation}
with atoms of mass $m$, scattering length $a$ and number density $n$. We introduce the dimensionless scaling function $b(t)$ so that the interaction strength (or equivalently the scattering length) has the time dependence
\begin{eqnarray}
U(t) \equiv U_0 \, b(t) ,
\end{eqnarray}
where $U_0 = U(t_0)$ at an initial time $t_0$; therefore we take $b(t_0)=1$. From (\ref{soundspeed}) the time dependence of the speed of sound is thus
\begin{equation}
c(t) =  c_0 \, b(t)^{1/2}.
\end{equation}
In practice a variation in the interaction strength is possible by using a Feshbach resonance \cite{Vogels1997, Inouye1998,Barcelo2003}. If $b(t)$ is decreasing with time we have an \emph{expanding} universe model, whereas if $b(t)$ increases with time we have a \emph{contracting} universe model.  It is useful to introduce $X$ the expansion of the universe between two times $t_0$ and $t_f$ in the following way:
\begin{equation} 
X = b(t_f)^{\alpha-1}.
\end{equation}
Please note that one can always without loss of generality choose $b(t_{0})=1$.
The line element for the FRW universe we have described is given by
\begin{equation} \label{FRWlineelement}
d s\eff^2 = \Omega^{2}_{0} \left[ -c_0^2 b(t)^{\alpha} \d t^2 +
b(t)^{\alpha- 1} \d \x^2 \right],
\end{equation}
where we have introduced the conformal factor
\begin{equation}
\Omega^{2}_{0}(n_0,c_0,d) = \left( \frac{n_0}{c_0}
\right)^{\frac{2}{d-1}},
\end{equation}
which is independent of space and time, as well as the dimension-dependent exponent
\begin{equation}
\alpha = \frac{d-2}{d-1}.
\end{equation}
Hence, in $d=2$ spatial dimensions we get 
\begin{equation} \label{X}
X = b(t_f)^{-1},
\end{equation}
as the overall expansion of the universe between $t_{0}$ and $t_{f}$.

In what respect do we have an expanding universe, given that the condensate is contained in a physically fixed volume $V$? A decrease in the scattering length corresponds to a decrease in the the speed of sound propagating in the condensate; therefore any acoustic excitations will propagate with decreasing speed in the condensate as time passes. To an observer at rest in the effective spacetime, a decrease of the speed of sound is thus indistinguishable to an isotropic expansion of the spatial dimensions. In this sense, (\ref{frw_metric}) corresponds to the notion of a spacetime undergoing an \emph{effective expansion}.

It is not straightforward to define either an \emph{apparent} or \emph{event} horizon in the model considered here as the system is homogeneous, and the background velocity is therefore the same (\emph{i.e.}, zero) everywhere. This is further complicated by the fact that the causal structure of the effective spacetime should be determined by the maximum \emph{signal} velocity (\emph{i.e.}, group velocity), which is effectively infinite here owing to the super-phononic modes in a BEC. Alternatively, analogue models where the background velocity depends on the radial position (\emph{e.g.}, when the trapping potential is switched off and the condensate is free to expand) have been studied in \cite{Weinfurtner2004,Fedichev2003,Fedichev2003a,Fedichev2004,Uhlmann2005}. In the present situation, where the scale factor $a_{\textrm{FRW}}(t)$ or equivalently $b(t)$ contains all the geometric structure for the spacetime, it is necessary to perform the usual analysis in terms of \emph{cosmological horizons} to determine the overall causal structure of the spacetime \cite{EllisRothman}.

%+++++++++++++++++++++++++++++++++++++++++++++++++++++++++++++++++
\subsection{Two dimensional model}
%+++++++++++++++++++++++++++++++++++++++++++++++++++++++++++++++++
To facilitate the numerical calculations required by the classical field simulations that we present in Sec.~\ref{chap:expandingsimulations}, we continue with $d = 2$ spatial dimensions. The reduced mode space for $d = 2$ greatly decreases the computation time, but still leads to a satisfactory description of the system. In particular, we expect the extension of the numerical simulations to $d = 3$ to lead to qualitatively similar results 
\footnote{This is because the projected Gross--Pitaevskii equation 
--- which the classical field method is based on --- takes 
the same form for $d = 2$ or $d=3$. While the extension of 
the classical field method to $d = 3$ spatial dimensions 
is straightforward, this requires a larger mode space 
and our simulations would have required an unfeasibly 
large computational time. }.
Moreover, while it has been shown \cite{Fedichev2004,Uhlmann2005} that for $d = 2$, a condensate undergoing free expansion leads to a scalar field equation that does not depend on the scaling factor in co-moving coordinates \footnote{The models considered in \cite{Fedichev2004,Uhlmann2005} correspond to the case of a \emph{physical expansion}, which corresponds to the free expansion of a BEC. The scaling factor there has a different physical significance from the scaling factor for the \emph{effective expansion} that is considered here.} --- so that in this case there can be no particle production --- as we will see, the present model for a $2+1$ dimensional FRW universe leads to a scalar field equation that \emph{does} allow particle production.

For $2+1$ spacetime dimensions, $\alpha(2)=0$ and the line-element simplifies to
\begin{equation}\label{met2Dconstrho}
d s\eff^2 = \left(\frac{n_0}{c_0}\right)^2 \left[ -c_0^2 \d t^2+
b(t)^{-1} \d\x^2 \right].
\end{equation}
A further time transformation is not required as laboratory time and proper time (for a comoving observer) are of the same form.
Although (\ref{met2Dconstrho}) includes a time- and space-independent conformal factor, this does not affect the dynamics of the field (in this case the mode functions need to be normalized for consistency with the Bogoliubov theory). 
The scaling factor $a_{\textrm{FRW}}(t)$ for a FRW universe that is familiar from cosmology is related to $b(t)$ by
\begin{eqnarray}
a_{\textrm{FRW}}(t) = b(t)^{-1/2}
\end{eqnarray}
for $d = 2$. (We always explicitly specify the FRW subscript for the cosmological scale factor so that this quantity is not confused with the $s$-wave scattering length $a$.)

The reduction of the model to two dimensions requires a modification to the nonlinearity that appears in the GPE (\ref{eq:gpe}), and therefore also the resulting field equation (\ref{fieldequation_f}). To see this, we assume the transverse $z$ dimension is tightly confined with the trap lengths satisfying $L_z \ll L_x$, $L_y$, and further that the $L_z \sim \xi$ for the transverse dimension where $\xi$ is the healing length of the condensate. The scattering is still determined by the three-dimensional scattering length so that this is called a \emph{quasi-two dimensional geometry}. However, the system remains in the ground state of the transverse dimension because the energy required for transverse excitations is much larger than for longitudinal excitations. The wavefunction is then separable as $\Psi(\mathbf{x}, t) = \psi(x,y,t) \zeta(z)$. Assuming the condensate is homogeneous in the $z$ direction, the GPE can be rewritten as
\begin{eqnarray}\label{eq:gpe2d}
i \hbar \frac{\partial \psi(x, y, t)}{\partial t} &=& \left (-\frac{\hbar^2}{2 m} \nabla^2 + V_{\textrm{ext}} \right ) \psi(x, y, t) \nonumber \\ 
&&+ U_{\textrm{2D}} |\psi(x, y, t)|^2 \psi(x, y, t),
\end{eqnarray}
where the effective nonlinearity is $U_{\textrm{2D}} = U/L_z$. This does not affect the form of the resulting calculations for particle production, but should be noted when we determine suitable parameters for our simulations in Sec.~\ref{sect:cfmsuitable}. (Note that because we are changing the atomic interactions by several orders of magnitude during the proposed experiment, this will also affect the transverse trapping.)

%%%%%%%%%%%%%%%%%%%%%%%%%%%%%%%%%%%%%%%%%%%%%%%%%%%
%
\section{Field quantization and particle production}\label{sect:quantisation}
%
%%%%%%%%%%%%%%%%%%%%%%%%%%%%%%%%%%%%%%%%%%%%%%%%%%%
The field $\otheta(t)$ is quantized using the plane wave mode expansion:
\begin{eqnarray}\label{mode_expansion}
\otheta(\mathbf{x}, t) = \frac{1}{\sqrt{V}} \sum_{\mathbf{k}} \left[ \obk e^{i \mathbf{k}\cdot\mathbf{x}} \chi_{\mathbf{k}}(t) + \obkdagg e^{- i \mathbf{k}\cdot\mathbf{x}} \chi^{*}_{\mathbf{k}}(t) \right],
\end{eqnarray}
where $\obk$ and $\obkdagg$ are the annihilation and creation operators respectively for the quasiparticle modes. 

In flat (Minkowski) spacetime we can associate the positive and negative frequency solutions of (\ref{fieldequation_g}) with annihilation and creation operators respectively. In curved spacetime, this association is not always possible as different observers experience different vacua; in the language of general relativity, the spacetime need not have a (globally defined) time-like Killing vector field so that the positive frequency solution is not necessarily unique and typically is not an eigenfunction of $\partial_t$. A consequence of this is that the ``natural'' choice of a Fock vacuum according to $\hat{b} |0\rangle = 0$ depends in general on the choice of coordinates; that is, the measurement of particle content in curved spacetime is said to be ``observer dependent''. 

The calculation of particle production follows the standard methodology \cite{Birrell1982,Fulling1989}. 
We define \emph{in} and \emph{out} regions respectively as asymptotically flat regions with $t \rightarrow -\infty$ and $t \rightarrow +\infty$. (While the existence of asymptotically flat regions cannot be assumed for cosmological models,  it is certainly possible to emulate this scenario in BEC experiments.) We can write a mode expansion (\ref{mode_expansion}) for the \emph{in} and \emph{out} regions in terms of mode functions $\chi_{\mathbf{k}}^{\inss}$ or $\chi_{\mathbf{k}}^{\outss}$ respectively. Because both sets of modes are a basis set for the field, they are related by the Bogoliubov transformation
\begin{eqnarray}\label{Bogmodemixing}
\chi_{\mathbf{k}}^{\outss} = \alpha_{\mathbf{k}} \chi_{\mathbf{k}}^{\inss} + \beta_{\mathbf{-k}} \chi_{\mathbf{-k}}^{\inss *}.
\end{eqnarray}
Note the spacetime has translational invariance as a symmetry so the field (\ref{mode_expansion}) can be expanded in the same set of plane wave modes $e^{i \mathbf{k} \cdot \mathbf{x}}$ for both the \emph{in} and \emph{out} regions, and therefore $\beta_{\mathbf{-k}} = \beta_{\mathbf{k}}$ also. By convention the mode functions are normalised according to the Klein--Gordon inner product. In momentum space this takes the form
\begin{eqnarray}\label{modesnorm}
 -i W[\chi_{\mathbf{k}},  \chi_{\mathbf{k}}^*] = 1,
\end{eqnarray}
where $W[f_1, f_2] = f_1 (\partial_{t} f_2) - (\partial_{t} f_1) f_2$ is the Wronskian of two functions $f_1$ and $f_2$. 

For each asymptotic region the vacuum state is defined by the requirement
\begin{eqnarray}
b^{\inss}_{\mathbf{k}} | 0 ^{\inss} \rangle = 0, \nonumber \\
b^{\outss}_{\mathbf{k}} | 0 ^{\outss} \rangle = 0.
\end{eqnarray}
The \emph{in} and \emph{out} mode operators are also related by
\begin{eqnarray}\label{Bogmodemixing2}
b^{\outss}_{\mathbf{k}} = \alpha_{\mathbf{k}} b^{\inss}_{\mathbf{k}} + \beta_{-\mathbf{k}} b^{\dagger {\inss}}_{-\mathbf{k}}.
\end{eqnarray}
The number of particles created is then given by the quantity
\begin{eqnarray}
N_{\mathbf{k}}^\outss & = & \langle 0 ^{\inss} | b^{\dagger {\outss}}_{\mathbf{k}} b^{\outss}_{\mathbf{k}} | 0 ^{\inss} \rangle \nonumber \\
& = & |\beta_{\mathbf{k}}|^2.
\end{eqnarray}
The standard procedure to calculate $\beta_{\mathbf{k}}$ involves solving the field equation (\ref{fieldequation_f}) for the \emph{in} and \emph{out} regions by using (\ref{Bogmodemixing}) with the appropriate boundary conditions. 

It should be stressed that here particle production refers to the production of (massless) quasiparticle excitations in the Bogoliubov basis, that approximately diagonalizes the many body Hamiltonian (\ref{secondquantH}) for the Bose gas. As (\ref{Bogmodemixing2}) couples modes of momenta $\mathbf{k}$ and $-\mathbf{k}$ this process is often referred to as \emph{pair production} and is associated with the formation of \emph{squeezed states} \cite{Grishchuk1990, Liberati2000,Jacobson2004,Liberati:1998tf,Belgiorno:1999ha}.

%~~~~~~~~~~~~~~~~~~~~~~~~~~~~~~~~~~~~~~~~~~~~~~~~~~~~
\subsubsection{Choosing a Fock vacuum}
%~~~~~~~~~~~~~~~~~~~~~~~~~~~~~~~~~~~~~~~~~~~~~~~~~~~~
When the \emph{in} and \emph{out} regions of the expansion are not asymptotically flat, a vacuum state cannot be unambiguously defined for either case. However, clearly the procedure to calculate particle production outlined above requires a choice of Fock vacuum, and moreover, that choice should lead to physically reasonable results. We therefore mention two choices that have been developed to deal with this situation (although there are many more):
\begin{itemize}
\item[] {\bf The instantaneous Minkowski vacuum:} this is the state that corresponds to the instantaneous diagonalization of the Hamiltonian at a given time (and is therefore also the state that minimises the energy). (This procedure is also known as \emph{Hamiltonian diagonalization}.) This choice may be problematic in some cases as it can lead to situations of infinite particle production even though the expansion may be smooth and finite \cite{Fulling1989}. 
\item[]  {\bf The adiabatic vacuum:} this approximation can be used for modes that experience a sufficiently slow expansion and is formulated in terms of the WKB approximation \cite{Birrell1982, Winitzki2005}.  However, the adiabaticity requirement means it has limited applicability. 
\end{itemize}
In Sec.~\ref{sect:desitteracoustic} we calculate particle production for the case of a de Sitter expansion, for which there are no static regions --- in this case we therefore utilize the instantaneous Minkowski vacuum to calculate the Bogoliubov coefficients at a given time during the expansion. Once the expansion has stopped the final Bogoliubov coefficients unambiguously correspond to physical particle production. The problem of infinite particle production does not occur in this case, and as we shall see from the classical field simulation results in Sec.~\ref{chap:expandingsimulations}, particle production is further suppressed for short wavelength modes because of super-phononic dispersion.

%~~~~~~~~~~~~~~~~~~~~~~~~~~~~~~~~~~~~~~~~~~~~~~~~~~~~
\subsubsection{Commutation relations}
%~~~~~~~~~~~~~~~~~~~~~~~~~~~~~~~~~~~~~~~~~~~~~~~~~~~~
In addition to identifying an effective metric required to make the analogy with quantum field theory in curved spacetime, another prerequisite is that in quantizing the field the operators should be annihilation and creation operators in the usual sense. That is, they should obey the correct commutation relations so that we can define quanta of the field in the Fock basis. This has indeed shown to be the case for the linear excitations of a BEC \cite{UnruhSchutzhold2003,Weinfurtner:2007aa}, and also follows from the discussion in Sec.~\ref{bogtheory}.

%~~~~~~~~~~~~~~~~~~~~~~~~~~~~~~~~~~~~~~~~~~~~~~~~~~~~
\subsubsection{Relevant scales}
%~~~~~~~~~~~~~~~~~~~~~~~~~~~~~~~~~~~~~~~~~~~~~~~~~~~~
There are two relevant scales for particle production in a FRW-type analogue model:
\begin{enumerate}
\item Hubble parameter - Following \cite{Barcelo2003} for the metric (\ref{met2Dconstrho}) the Hubble parameter is given, for $d=2$, in terms of the scaling function $b(t)$ in laboratory time by
\begin{eqnarray}\label{hubble}
H \equiv \frac{\dot{a}_{\textrm{FRW}}}{a_{\textrm{FRW}}}=  -\frac{1}{2} \, \frac{\dot{b}}{b}.
\end{eqnarray}
This quantity, which is time-dependent in general, corresponds to the rate of expansion for the universe. If $H \gtrsim \omega_k$ for a mode frequency labeled by $k$ the dynamics are \emph{Hubble-dominated} and we expect the mode to be to be non-oscillating, whereas $H \ll \omega_k$ implies the mode is oscillating and in the adiabatic regime. In the latter case this is what cosmologists refer to as a \emph{parametrically excited} mode. 
Clearly a large value for $H$ is favourable to particle production --- however $H$ can not be made arbitrarily large as the approximations that lead to the effective field theory would then be violated \cite{Barcelo2003,Fischer2004}.

\item Healing length -  In a BEC, the healing length is a distance over which localized perturbations in the condensate tend to smooth out, and is given by 
\begin{eqnarray}\label{healinglength}
\xi = \frac{\hbar}{\sqrt{2} m c} = \xi_0 \, b(t)^{-1/2}.
\end{eqnarray}
That is, if we define the cross-over from phonon to free-particle behaviour for the Bogoliubov spectrum  (\ref{bogdispersion}) as $\hbar^2 k_c^2/2 m \equiv U n$ then $k_c = 1/\xi$. Modes for which $k \ll 1/\xi$ correspond to collective excitations of the condensate (phonons) and couple to the effective time-dependent curved spacetime whereas modes with $k \gg 1/\xi$ are particle-like and are relatively unaffected by the emergent spacetime geometry. Alternatively stated, the spacetime appears locally flat and time-independent to modes with sufficiently short wavelengths. 

\end{enumerate}

Particle production into a mode $\mathbf{k}$ then proceeds as follows: During expansion, while $H \gtrsim \omega_k$ and $k < 1/\xi$, the mode evolves non-trivially and particle production will certainly occur. As the expansion proceeds, particle production can ``switch off'' for two reasons: Either the healing length increases until the mode becomes particle-like and particle production slows, ceasing altogether when $k \gg 1/\xi$; or alternatively, if the expansion slows, then after some time $H \lesssim \omega_k$ so that the field begins to oscillate relatively freely, and additional particle production also ceases. In either case, the mode no longer evolves, and the occupation number of the mode becomes constant.

In general, it can be difficult or impossible to analytically solve the field equation (\ref{fieldequation_f}) for the mode functions. Moreover, if the \emph{in}\,/\,\emph{out} region is not flat, there is no preferred choice for the initial\,/\,final vacuum state. However, when the field evolution is sufficiently slow --- \emph{i.e.}, adiabatic --- the WKB approximation can be used to calculate particle production (see \cite{Winitzki2005} and references therein). It is worth noting that this approach leads to a Planckian spectrum in the lowest order approximation \cite{Birrell1982}. 

%~~~~~~~~~~~~~~~~~~~~~~~~~~~~~~~~~~~~~~~~~~~~~~~~~~~~
\subsubsection{A note on freezing}
%~~~~~~~~~~~~~~~~~~~~~~~~~~~~~~~~~~~~~~~~~~~~~~~~~~~~
For our analogue model of an expanding FRW universe, in the acoustic approximation we have $\omega_k(t) = c(t) k$, which decreases as the expansion proceeds (this is equivalent to the usual notion of \emph{cosmological redshifting} of modes that occurs during inflation). This means that depending on the form of the expansion, a mode that is initially oscillating ($H \ll \omega_k$) may enter a Hubble-dominated era ($H \gg \omega_k$) after a sufficiently long expansion --- this is the mechanism for \emph{freezing} of modes that is familiar from inflationary cosmology. The term \emph{freezing} does not necessarily imply that particle production ceases, but rather, that the mode no longer oscillates. 

The situation is quite different when nonlinear super-phononic dispersion is included --- in this case, the healing length $\xi$ also decreases for increasing time, which means that a mode that was initially phononic will crossover into the adiabatic free-particle regime for a sufficiently long expansion. In this case therefore, the notion of \emph{freezing} can only occur in a transient regime that depends on the mode wave-vector $k$ and the form of the scaling function $b(t)$ (and therefore on the Hubble parameter also). A more detailed discussion can be found in \cite{Weinfurtner:2007ab}.

%+++++++++++++++++++++++++++++++++++++++++++++++++++++++++++++++++
\subsection{Acoustic approximation}\label{acoustic2}
%+++++++++++++++++++++++++++++++++++++++++++++++++++++++++++++++++
In the acoustic approximation, the correspondence between the massless minimally coupled scalar field equation for a FRW universe and the equations of motion for linearized fluctuations in the BEC is exact. The dynamics of the low momentum modes are expected to fall into this regime, and therefore it is of interest to examine this case first.

%~~~~~~~~~~~~~~~~~~~~~~~~~~~~~~~~~~~~~~~~~~~~~~~~~~~~
\subsubsection{Conformal time}
%~~~~~~~~~~~~~~~~~~~~~~~~~~~~~~~~~~~~~~~~~~~~~~~~~~~~
We define a coordinate transformation from laboratory time $t$ to conformal time $\eta$ by
\begin{equation}
\d\eta = \sqrt{b(t)} \d t.
\end{equation}
The line element (\ref{met2Dconstrho}) then reads
\begin{equation}
ds\eff^2 = \Omega_{0}^{2} \; b(\eta)^{-1}  \left[ - c_{0}^{2} \d\eta^{2} + \d\mathbf{x}^{2} \right],
\end{equation}

In $(2+1)$ dimensions the equation of motion for the field obtained from (\ref{fieldequation_f}) in conformal time becomes
\begin{equation}\label{fieldequation_acoustic}
\partial^2_{\eta} \otheta  - \frac{1}{2} \frac{\dot{b}(\eta)}{b(\eta)} \partial_{\eta} \otheta - c_0^2 \, \nabla^2 \otheta  = 0.
\end{equation}
Note that the coefficient of $\partial_{\eta} \otheta$ is the Hubble parameter given by  (\ref{hubble}). 
The calculation of particle production requires the solution of this field equation for the mode functions $\chi(\eta)$. This task is assisted by reducing the field equation to standard form where the first order derivative of the field operator does not appear; 
this can be achieved (for instance) by introducing an auxiliary field of the form $\hat{\varsigma}(\eta) = b(\eta) \otheta$, showing mathematically equivalent dynamics to $\otheta$. Alternatively we can consider the transformation to \emph{auxiliary time} as follows.

%~~~~~~~~~~~~~~~~~~~~~~~~~~~~~~~~~~~~~~~~~~~~~~~~~~~~
\subsubsection{Auxiliary time}
%~~~~~~~~~~~~~~~~~~~~~~~~~~~~~~~~~~~~~~~~~~~~~~~~~~~~
Defining the auxiliary factor  by
\begin{eqnarray}
\Lambda = \Omega_0^{2/3} b(t)^{-1/3},  
\end{eqnarray}
and the \emph{auxiliary time} by
\begin{eqnarray}
\taux = \int \frac{c_0 b(t)}{\Omega_0} \; dt,
\end{eqnarray}
the line element (\ref{met2Dconstrho}) transforms as
\begin{eqnarray}
dl\eff^2 = -\Lambda^6 d\taux^2 + \Lambda^3 d\mathbf{x}^2.
\end{eqnarray}
For a massless scalar field, the corresponding field equation in $(2+1)$ dimensions is
\begin{eqnarray}
\partial^2_{\taux} \otheta - \Lambda^3 \nabla^2 \otheta = 0.
\end{eqnarray}
Using the field mode expansion given by (\ref{mode_expansion}) we then get a time-dependent harmonic oscillator for each mode
\begin{eqnarray}\label{fieldequation_k_auxiliary}
\partial^2_{\taux} \chi_{\mathbf{k}} +  \waux^2_{\mathbf{k}}(\taux) \chi_{\mathbf{k}} = 0
\end{eqnarray}
where $\waux^2_{\mathbf{k}}(\taux) = \Lambda^3 k^2 = \Omega_0^2 b(\taux)^{-1} k^2$
is the oscillator frequency.
While the \emph{auxiliary time} approach is useful in the acoustic approximation, it does not generally lead to the simple form (\ref{fieldequation_k_auxiliary}) when the quantum pressure term is included (see Sec.~\ref{lorentzbreak}).

%+++++++++++++++++++++++++++++++++++++++++++++++++++++++++++++++++
\subsection{Beyond the acoustic approximation}\label{lorentzbreak}
%+++++++++++++++++++++++++++++++++++++++++++++++++++++++++++++++++
The description from the previous section was given within the acoustic approximation, valid only for long wavelength modes, whereby the quantum pressure term is omitted from the field equation (\ref{fieldequation_f}) and the resulting equations for the FRW analogue model given by (\ref{fieldequation_acoustic}) or (\ref{fieldequation_k_auxiliary}). We can extend the analysis to higher momentum modes by including the quantum pressure term; this leads to a super-phononic type dispersion relation for high frequency modes and an appropriately modified field equation as we presently discuss. Moreover, this description corresponds to the presence of ``trans-Planckian'' like modes, and leads to the idea of Lorentz violation.  Such \emph{modified dispersion relations} have similarly been incorporated into some studies of inflationary cosmology --- we discuss this here briefly, but dedicate our companion paper \cite{Weinfurtner:2007ab} to exploring this subject. In Sec.~\ref{sect:nonlindispresults} we will consider two specific forms of the scaling function in this regime: the limiting case of a sudden transition (Sec.~\ref{sect:suddenfull}) and a cyclic universe model (Sec.~\ref{sect:cyclicacoustic}), both which include the effect of quantum pressure. 

%~~~~~~~~~~~~~~~~~~~~~~~~~~~~~~~~~~~~~~~~~~~~~~~~~~~~
\subsubsection{Nonlinear dispersion}\label{sect:nonlindisp}
%~~~~~~~~~~~~~~~~~~~~~~~~~~~~~~~~~~~~~~~~~~~~~~~~~~~~
We include the quantum pressure term by using (\ref{nkgeneral}) so that the field equation (\ref{fieldequation_f}) in momentum space becomes
\begin{eqnarray}\label{fieldequation_nonlin}
\partial^2_{t} \chi_k - \frac{1}{2} \frac{c_0^2 k^2}{\omega_k(t)^2} \, \partial_{t} b(t) \,
 \partial_{t} \chi_k + \omega_k(t)^2 \, \chi_k = 0,
\end{eqnarray}
where we have now defined
\begin{eqnarray}\label{omega_nonlin}
\omega_k(t)^2 = \frac{k^2}{2 m} \left(  \frac{\hbar^2 k^2}{2 m} + 2 U(t) n_0 \right).
\end{eqnarray}
It should come as no surprise that we have recovered the Bogoliubov dispersion relation for a weakly interacting Bose gas (see Sec.~\ref{sect:bogconnection} below). In particular, we see that if we set $b(t) = \textrm{const.}$ then the field equation describes the dynamics of each Bogoliubov mode for a time-independent Hamiltonian.

In general, it is difficult to solve (\ref{fieldequation_nonlin}) for the mode functions, and to hence calculate particle production. This situation often persists even though it may be possible to reduce (\ref{fieldequation_nonlin}) to standard form where the first order derivative term $\partial_{t} \theta_k$ does not appear. We can, however, make a qualitative statement about particle production for high momenta modes by considering the crossover from phononic to free-particle modes as determined by the healing length (\ref{healinglength}). For modes that satisfy $k \gg k_c$, the field equation takes the approximate form
\begin{eqnarray}\label{fieldequation_nonlin_largek}
 \partial^2_{t} \chi_k  + \left(\frac{\epsilon_k^0}{\hbar}\right)^2 \chi_k = 0,
 \end{eqnarray}
 where
\begin{equation}
\label{Eq:Single.Particle.Energy}
\epsilon_{k}^{0} = \frac{\hbar^{2} \, k^{2}}{2\, m}
\end{equation}
represents the single particle energy for a non-interacting gas.
That is, the time dependence from $b(t)$ above is largely suppressed, and each mode with $k \gg k_c$ evolves trivially as a time-independent harmonic oscillator, remaining in its initial vacuum state. Therefore these modes experience \emph{no} particle production. 

\subsubsection{Lorentz violation}\label{sect:frwlorentz}

Planck-scale Lorentz violation is a feature of some theories of quantum gravity that can be modeled by the presence of a \emph{modified dispersion relation} at the Planck scale \cite{Mattingly2005,Liberati:2005id,Weinfurtner:2006nl,Weinfurtner:2006iv,Weinfurtner:2006eq,Liberati:2006sj}. In this sense, for a BEC the healing length $\xi$ provides the analogue of the Planck scale, characterising the cross-over from phononic ($k \ll 1/\xi$) to free-particle ($k \gg 1/\xi$) like modes. Specifically, the field equation (\ref{fieldequation_nonlin}) is not Lorentz invariant because of the nonlinear super-phononic dispersion of the modes (\ref{omega_nonlin}) at large momenta (this clearly evident from Eq. (\ref{fieldequation_nonlin_largek}) where the second term varies as $\sim k^4$). On the other hand, for small momenta (\emph{i.e.}, in the acoustic approximation) the dispersion relation is linear in $k$ and the field equation reduces to the Lorentz invariant form (\ref{fieldequation_acoustic}). It should be stressed however, that the effective quantum field theory for the BEC is still valid as long as atomic interactions can be characterised by the $s$-wave scattering length, which is true in general for some cut-off in wave-vector $k_{\textrm{cut-off}} > 1/\xi$. 

We note that modified dispersion relations have sometimes been used to study the ``trans-Planckian'' problem in inflationary cosmology \cite{Martin2001,Kempf:2006al,Mersini-Houghton:2001aa,Bastero-Gil:2003aa} --- the results there showed that certain modifications to the dispersion relation could lead to significant deviations for the density fluctuation spectrum when compared to the unmodified dispersion relation (\emph{i.e.}, the usual model of inflation).

%----------------------------------------------------------------------------
\section{Connection with Bogoliubov Theory}\label{sect:bogconnection}
%----------------------------------------------------------------------------

Thus far, we have derived a field equation for a scalar field propagating in an effective spacetime. The quanta of this scalar field must correspond to the linearised quantised excitations (\emph{i.e.}, phonons) of the quantum field for the theory to be consistent. Therefore, at this point, to make this connection explicit, it is worth reviewing the theory of quantum excitations in BECs, which is well described by Bogoliubov's theory of excitations for a weakly interacting system.

%----------------------------------------------------------------------------
\subsection{Bogoliubov Theory}\label{bogtheory}
%----------------------------------------------------------------------------

While the GPE has been extremely successful for describing mean-field effects (\emph{i.e.}, classical dynamics), it neglects quantum and thermal fluctuations.  To first order, we can include quantum fluctuations in this description by considering the theory of elementary excitations for a weakly interacting Bose gas, first formulated by Bogoliubov for the homogeneous case \cite{Bogoliubov1947}.
The standard procedure is to first expand  the field operator in a plane-wave basis
\begin{eqnarray}\label{psiexpansion}
\hat{\psi}(\x, t) = \frac{1}{\sqrt{V}} \sum_{\mathbf{k}} e^{i \mathbf{k} \cdot \mathbf{x}} \hat{a}_{\mathbf{k}} (t).
\end{eqnarray}
If the number of atoms in the condensate is large (\emph{i.e.}, the field is highly condensed) then replacing $\hat{a}_0$, $\hat{a}_0^{\dagger} \rightarrow \sqrt{N_0}$ and retaining terms of at least order $N_0$, the Hamiltonian (\ref{secondquantH}) can be approximately diagonalised using the Bogoliubov transformation
\begin{eqnarray}\label{bog_transform}
\hat{a}_{\mathbf{k}} &=& u_{\mathbf{k}} \hat{b}_{\mathbf{k}} + v_{\mathbf{k}} \hat{b}_{-\mathbf{k}}^{\dagger}, \nonumber \\
\hat{a}_{-\mathbf{k}}^{\dagger} &=& u_{\mathbf{k}} \hat{b}_{-\mathbf{k}}^{\dagger} + v_{\mathbf{k}} \hat{b}_{\mathbf{k}}, 
\end{eqnarray}
when $u_{\mathbf{k}}$ and $v_{\mathbf{k}}$ are chosen so that the new operators $\hat{b}_{\mathbf{k}}$ and $\hat{b}_{\mathbf{k}}^{\dagger}$ satisfy the commutation relations for Bose field operators
\begin{eqnarray}
u_{\mathbf{k}}^2 - v_{\mathbf{k}}^2 = 1.
\end{eqnarray}
This leads to the result that \cite{FetterWalecka, Abrikosov1963}
\begin{eqnarray}\label{bogukvk}
u_{\mathbf{k}} = \frac{1}{\sqrt{1 - A_{\mathbf{k}}^2}} \textrm{,} \hspace{0.5cm} 
v_{\mathbf{k}} = \frac{A_{\mathbf{k}}}{\sqrt{1 - A_{\mathbf{k}}^2}},
\end{eqnarray}
with
\begin{eqnarray}\label{bogAk}
A_{\mathbf{k}} = \frac{1}{U_0 n} \left[\,- (\epsilon_{\mathbf{k}}^0 + U_0 n) + \sqrt{\epsilon_{\mathbf{k}}^0 (\epsilon_{\mathbf{k}}^0 + 2 U_0 n)}\,\right],
\end{eqnarray}
where $n = N_0/V$ is the density. The resulting Hamiltonian is
\begin{eqnarray}\label{eq:Hbog}
\hat{H} \approx \hat{H}_{\scriptsize\textrm{Bog}} = E_0 + \sum_{\mathbf{k} \neq 0} \epsilon_{\mathbf{k}} \hat{b}_{\mathbf{k}}^{\dagger} \hat{b}_{\mathbf{k}},
\end{eqnarray}
with a constant $E_0$ and where the quasiparticle excitations have the energy spectrum
\begin{eqnarray}\label{bogdispersion}
\epsilon_{\mathbf{k}} = \sqrt{\epsilon_{\mathbf{k}}^0 (\epsilon_{\mathbf{k}}^0 + 2 U_0 n)},
\end{eqnarray} 
in terms of the single particle energy for a non-interacting gas, see Eq.~\refb{Eq:Single.Particle.Energy}.
In the Bogoliubov approximation, explicitly pulling out a phase factor depending on the chemical potential $\mu = U n_0$. the Bose field operator can conveniently be expanded as
\begin{eqnarray}
\hat{\psi}(\mathbf{x}, t) & = &  e^{-i \mu t/\hbar} 
\left[\sqrt{n_0} + \delta \hat{\varphi}(\mathbf{x}, t) \right]  ,
\end{eqnarray}
where
\begin{equation}
\delta \hat{\varphi}(\mathbf{x}, t) = \sum_k \left[ U_k(\mathbf{x}, t) \hat{b}_{\mathbf{k}}(0)  + V_k^*(\mathbf{x}, t) \hat{b}^{\dagger}_{\mathbf{k}}(0) \right].
\end{equation}
Note that the time-dependence of each mode is fully contained in the mode functions $U_k(\mathbf{x}, t)$ and $V_k(\mathbf{x}, t)$. When the Hamiltonian is time-independent (\emph{i.e.}, $U$ constant) the time-dependence is purely oscillatory and the number of quasiparticles in each mode $\langle \hat{b}_{\mathbf{k}}^{\dagger} \hat{b}_{\mathbf{k}} \rangle$ is a constant of the motion. 

Alternatively when we linearize the density-phase representation of the field operator we find
\begin{eqnarray}
\delta \hat{\varphi}(\mathbf{x}, t) \approx \sqrt{n_0} \left( \frac{\hat{n}_1}{2 n_0} + i \hat{\theta}_1 \right).
\end{eqnarray}
If $\hat{n}_1$ and $\hat{\theta}_1$ are Hermitian we can write
\begin{eqnarray}
\hat{n}_1 = \sqrt{n_0} ( \delta \hat{\varphi} + \delta \hat{\varphi}^{\dagger} ),
\end{eqnarray}
\begin{eqnarray}
\hat{\theta}_1 = \frac{1}{2 \sqrt{n_0} i} ( \delta \hat{\varphi} - \delta \hat{\varphi}^{\dagger} ).
\end{eqnarray}
The commutation relations for the field operator (\ref{psicommutation}) can be used to show that the density and phase fluctuations satisfy the commutation relation
\begin{eqnarray}\label{ntheta_commutator}
[\hat{n}_1(\mathbf{x}), \hat{\theta}_1(\mathbf{x}^{\prime}) ] = i \delta(\mathbf{x} - \mathbf{x}^{\prime}).
\end{eqnarray}

Using the mode expansion (\ref{mode_expansion}) we expand the field $\hat \theta_1$ in Fourier (plane-wave) modes as
\begin{eqnarray}\label{theta1_expansion}
\hat{\theta}_1 = \frac{1}{\sqrt{V}} \sum_{\mathbf{k}} [ e^{i \mathbf{k} \cdot \mathbf{x}} \chi_{\mathbf{k}} \hat{b}_{\mathbf{k}}(t) +  e^{- i \mathbf{k} \cdot \mathbf{x}} \chi_{\mathbf{k}}^* \hat{b}_{\mathbf{k}}^{\dagger}(t)],
\end{eqnarray}
and similarly for $\hat n_1$
\begin{eqnarray}\label{n1_expansion}
\hat{n}_1 = \frac{1}{\sqrt{V}} \sum_{\mathbf{k}} [ e^{i \mathbf{k} \cdot \mathbf{x}} n_{\mathbf{k}} \hat{b}_{\mathbf{k}}(t) +  e^{- i \mathbf{k} \cdot \mathbf{x}} n_{\mathbf{k}}^* \hat{b}_{\mathbf{k}}^{\dagger}(t)].
\end{eqnarray}

The commutation relation (\ref{ntheta_commutator}) thus reduces to:
\begin{eqnarray}\label{wronskiancommutator}
n_k \chi_k^* - n_k^* \chi_k = i.
\end{eqnarray}

We can expand $\delta \hat{\varphi}$, $\hat{n}_1$ and $\hat{\theta}_1$ in the same set of plane wave modes, using the mode expansions (\ref{theta1_expansion}) and (\ref{n1_expansion}) and 
\begin{equation}
U_k(\mathbf{x}, t) =  u_k(t) \, e^{i \mathbf{k} \cdot \mathbf{x}}/\sqrt{V}
\end{equation}
and 
\begin{equation}
V_k(\mathbf{x}, t) =  v_k(t) \, e^{i \mathbf{k} \cdot \mathbf{x}}/\sqrt{V}, 
\end{equation}
so that the Fourier components are
\begin{eqnarray}\label{ukt}
u_k(t) = \frac{1}{2 \sqrt{n_0}} \, n_k(t) + i \sqrt{n_0} \, \chi_k(t),
\end{eqnarray}
\begin{eqnarray}\label{vkt}
v_k(t) = \frac{1}{2 \sqrt{n_0}} \, n_k(t) - i \sqrt{n_0} \, \chi_k(t).
\end{eqnarray}
Clearly these mode functions are consistent with the requirement that the Bogoliubov modes are normalised by $|u_k|^2 - |v_k|^2 = 1$. 
We note that these are general expressions, which are valid within the linearized theory of excitations regardless of what form the mode functions take. That is, they are valid for arbitrary forms of the scaling function $b(t)$ so long as the mode functions $n_k(t)$ and $\chi_k(t)$ can be found. 

%----------------------------------------------------------------------------
\subsection{Bogoliubov modes --- Minkowski spacetime}
%----------------------------------------------------------------------------

For the homogeneous model presented here the differential term (\ref{D_2}), accounting for the quantum pressure term,  takes the simple form
\begin{eqnarray}
\Dtwo \hat n_1 = \frac{1}{2 n_0} \nabla^2 \hat n_1.
\end{eqnarray}
Rearranging (\ref{theta1_linearized}) and extracting the Fourier component then gives
\begin{eqnarray}\label{nkgeneral}
n_{\mathbf{k}} = - \frac{\hbar}{U} \left[\frac{\epsilon_k^0}{2 U n_0} + 1 \right]^{-1} 
\partial_t \chi_{\mathbf{k}}.
\end{eqnarray}
Using (\ref{nkgeneral}) with the mode expansions (\ref{theta1_expansion}) and (\ref{n1_expansion}), the commutation relation (\ref{ntheta_commutator}) is satisfied when the mode functions take the form
\begin{eqnarray}\label{modefunctions}
\chi_{\mathbf{k}} = \frac{1}{2 \sqrt{n_0}} \sqrt{\frac{\epsilon_k}{\epsilon_k^0}} e^{- i \omega_{\mathbf{k}} t},
\nonumber \\
n_{\mathbf{k}} = i \sqrt{n_0} \sqrt{\frac{\epsilon_k^0}{\epsilon_k}} e^{- i \omega_{\mathbf{k}} t}.
\end{eqnarray}

These are the positive-frequency solutions for a time-independent Hamiltonian --- that is, when when $U$ is constant --- and correspond to an effective spacetime geometry that is Minkowski flat. The mode functions (\ref{ukt}) and (\ref{vkt}) can therefore be written
\begin{eqnarray}
|u_k| = \half \left\vert \sqrt{\frac{\epsilon_k}{\epsilon_k^0}} + \sqrt{\frac{\epsilon_k^0}{\epsilon_k}} \right\vert, \\
|v_k| = \half \left\vert \sqrt{\frac{\epsilon_k}{\epsilon_k^0}} - \sqrt{\frac{\epsilon_k^0}{\epsilon_k}} \right\vert.
\end{eqnarray}

%----------------------------------------------------------------------------
\subsection{Acoustic modes --- Minkowski spacetime}
%----------------------------------------------------------------------------
To facilitate the computation of particle production we again consider the acoustic approximation, which is valid for low momenta when $\hbar^2 k^2/2m \ll U n_0$ so that the quantum pressure term can be neglected. In this case (\ref{nkgeneral}) reduces to 
\begin{eqnarray}\label{nkacoustic}
n_{\mathbf{k}} = - \frac{\hbar}{U} \partial_t \chi_{\mathbf{k}}
\end{eqnarray}
The mode functions are then given by 
\begin{eqnarray}\label{chik}
\chi_k = \sqrt{\frac{U}{2 \hbar \omega_k}} e^{- i \omega_k t},
\end{eqnarray}
\begin{eqnarray}\label{nk}
n_k = i \sqrt{\frac{\hbar \omega_k}{2 U}} e^{- i \omega_k t}.
\end{eqnarray}
Using the general expressions (\ref{ukt}) and (\ref{vkt}) the mode functions are given by
\begin{eqnarray}
|u_k| = \frac{1}{2} \left\vert \sqrt{\frac{\hbar \omega_k}{2 U n_0}} + \sqrt{\frac{2 U n_0}{\hbar \omega_k}} \right\vert, \\
|v_k| = \frac{1}{2} \left\vert \sqrt{\frac{\hbar \omega_k}{2 U n_0}} - \sqrt{\frac{2 U n_0}{\hbar \omega_k}} \right\vert.
\end{eqnarray}

%----------------------------------------------------------------------------
\subsection{Quasiparticle production}\label{sect:bogproduction}
%----------------------------------------------------------------------------
The concept of particle production in our analogue model can be made explicit in terms of the Bogoliubov theory outlined above. To calculate the quasiparticle number in each mode after a finite expansion of duration $t$ it is necessary to project onto the Bogoliubov basis that instantaneously diagonalizes the many-body Bose Hamiltonian at $t$. 
(That is, we take the instantaneous Minkowski vacuum as the zero particle state.)
In what follows we take the initial condition as the quasiparticle vacuum at $t = 0$ so that $\hat{b}_{\mathbf{k}}(0) |0 \rangle = 0$ and therefore $N_k(0) = 0$. Here the mode functions are determined by the solutions to the time-independent case with $U(t) = U_0$, which yields the mode functions for a Minkowski spacetime. 
With this premise, we now proceed to calculate the particle production for each mode for a given expansion.
The Bogoliubov theory of a weakly interacting Bose gas predicts a non-zero depletion even at zero temperature; the \emph{real particle} annihilation operator is given by the time-dependent canonical transformation
\begin{eqnarray}
\hat{a}_{\mathbf{k}}(t) = u_k^{\expanding}(t) \, \hat{b}_{\mathbf{k}}(0) + v_k^{\expanding *}(t) \, \hat{b}_{-\mathbf{k}}^{\dagger}(0)
\end{eqnarray}
where $u_k^{\expanding}$ and $v_k^{\expanding}$ are solutions for the mode functions during the expansion --- these must coincide with Minkowski mode functions at $t = 0$ with $U(t) = U_0$. 
The projection into the Bogoliubov basis at $t$ is
\begin{eqnarray}
\hat{b}_{\mathbf{k}}(t) = u_k^{\outss *}(t) \,  \hat{a}_{\mathbf{k}}(t) - v_k^{\outss *}(t) \, \hat{a}_{-\mathbf{k}}^{\dagger}(t)
\end{eqnarray}
where $u_k^{\outss}$ and $u_k^{\outss}$ are given using (\ref{ukt}) and (\ref{vkt}) and the Minkowski mode solutions (\ref{chik}) and (\ref{nk}) with $U(t) = U_0/X$ in terms of the expansion $X$.
The particle production in each mode at time $t$ is then given by
\begin{eqnarray}\label{Nkbog}
N_k(t) & = &  \langle \hat{b}_{\mathbf{k}}^{\dagger}(t) \, \hat{b}_{\mathbf{k}}(t) \rangle \nonumber \\
& = & |u_k^{\outss *}(t) v_k^{\expanding *}(t)\!-\!v_k^{\outss *}(t) u_k^{\expanding *}(t)|^2 \nonumber \\
& = & (|u_k^{\outss}(t)|^2\!+\!|v_k^{\outss}(t)|^2)  \nonumber \\ 
&& \times \left[ \frac{1}{4 n_0} |n_k^{\expanding}(t)|^2\!+\!n_0 |\chi_k^{\expanding}(t)|^2 \right] 
\nonumber \\
&& - 2 |u_k^{\outss}(t)| |v_k^{\outss}(t)| \nonumber \\ 
&& \times \left[ \frac{1}{4 n_0} |n_k^{\expanding}(t)|^2 - n_0 |\chi_k^{\expanding}(t)|^2 \right] - \frac{1}{2}.
\end{eqnarray}
where we have used (\ref{ukt}) and (\ref{vkt}). Evidently, the central task is to solve the field equation (\ref{fieldequation_g}) for the mode functions $u_k^{\expanding}$ and $v_k^{\expanding}$ for a given expansion $b(t)$. This procedure will be applied in Sec.~\ref{sect:desitteracoustic} to calculate the quasiparticle production for the case of de Sitter expansion (in the acoustic approximation), and in Sec.~\ref{sect:suddenfull} for the case of sudden expansion (with high-frequency dispersion).

%%%%%%%%%%%%%%%%%%%%%%%%%%%%%%%%%%%%%%%%%%%%%%%%%%%
%
\section{Quasiparticle production in acoustic approximation}\label{sect:acousticpredictions}
%
%%%%%%%%%%%%%%%%%%%%%%%%%%%%%%%%%%%%%%%%%%%%%%%%%%%
We presently provide analytic solutions for quasiparticle production in the acoustic approximation for three different expansion scenarios: (i) Sudden transition; (ii) $\tanh$ expansion; and (iii) de Sitter expansion. Scenarios (i) and (ii) both have asymptoticly flat \emph{in} and \emph{out} regions, so the calculation of particle production follows the standard procedure from section \ref{sect:quantisation}. Scenario (iii) however does not have asymptoticly flat \emph{in} and \emph{out} regions for a finite time expansion, and therefore we must resort to the Bogoliubov theory of the previous section to calculate the particle production. 

%+++++++++++++++++++++++++++++++++++++++++++++++++++++++++++++++++
\subsection{Sudden transition $($$2+1$ dimensions$)$}\label{sect:suddenacoustic}
%+++++++++++++++++++++++++++++++++++++++++++++++++++++++++++++++++
A simple example of particle production is given by the limiting case of a sudden expansion. Here, the interaction is instantaneously switched from $U$ to $U/X$ at some time $\taux_0$ which we take as $\taux_0 = 0$ for convenience. The scaling function is given by
\begin{eqnarray}\label{bsudden}
b(\taux) = 1 - \left(1 - \frac{1}{X}\right) H(\taux)
\end{eqnarray}
where $H$ is the Heaviside step function. 
The case of particle production for a sudden transition has been previously explored by Jacobson for a parametric oscillator \cite{Jacobson2004}. 

Using the normalisation condition (\ref{modesnorm}) the positive frequency solutions to (\ref{fieldequation_k_auxiliary}) are given by
\begin{eqnarray}
\chi_k^{\inss/\outss} = \frac{1}{\sqrt{2 \waux_{\mathbf{k}}^{\inss/\outss}}} e^{- i \waux_{\mathbf{k}}^{\inss/\outss} \taux}.
\end{eqnarray}
It is straightforward to calculate the Bogoliubov coefficients by applying the boundary conditions from (\ref{Bogmodemixing}) and its first derivative for the Minkowski in and out modes at $t_0 = 0$. We find
\begin{eqnarray}
\alpha_{\mathbf{k}} = \half \left ( 
\sqrt{\frac{\waux_{\mathbf{k}}^{\inss}}{\waux_{\mathbf{k}}^{\outss}}}
+  \sqrt{\frac{\waux_{\mathbf{k}}^{\outss}}{\waux_{\mathbf{k}}^{\inss}}}
\right ),
\end{eqnarray}
and
\begin{eqnarray}
\beta_{\mathbf{k}} = \half \left ( 
\sqrt{\frac{\waux_{\mathbf{k}}^{\outss}}{\waux_{\mathbf{k}}^{\inss}}}
-\sqrt{\frac{\waux_{\mathbf{k}}^{\inss}}{\waux_{\mathbf{k}}^{\outss}}}
\right ),
\end{eqnarray}
so that
\begin{eqnarray}
\label{Nout_sudden}
N_{\mathbf{k}}^\outss = |\beta_{\mathbf{k}}|^2 = \quarter \left ( 
\sqrt{\frac{\waux_{\mathbf{k}}^{\inss}}{\waux_{\mathbf{k}}^{\outss}}}
-  \sqrt{\frac{\waux_{\mathbf{k}}^{\outss}}{\waux_{\mathbf{k}}^{\inss}}}
\right )^2.
\end{eqnarray}
Further noting for an expansion $X$ that
$\waux_{\mathbf{k}}^{\outss}/\waux_{\mathbf{k}}^{\inss} = 1/\sqrt{X}$, 
the particle production is then
\begin{eqnarray}\label{sudden_hydro}
N_{\mathbf{k}}^\outss = \quarter \left( X^{1/4} - X^{-1/4} \right )^{2}.
\end{eqnarray}
This yields, for example:  $N_{\mathbf{k}}^\outss \approx 0.4$ for $X = 10$;  $N_{\mathbf{k}}^\outss \approx 2$ for $X = 100$; and $N_{\mathbf{k}}^\outss \approx 11$ for $X = 2000$.

Equation (\ref{sudden_hydro}) provides an upper limit for the particle production in each mode \cite{Visser:1998ke}. This quantity does not depend on mode number --- a feature that reflects the fact that all modes experience a sudden change in the effective spacetime geometry. We shall see in section \ref{lorentzbreak} that including the quantum pressure term (\emph{i.e.}, nonlinear dispersion) in our formulation leads to suppressed particle production for increasing $|\mathbf{k}|$. 

It should be noted that the sudden transition corresponds to a delta function for the Hubble parameter $H$ at $\taux = 0$; this is physically unfeasible as any change in the $s$-wave scattering length via a Feshbach resonance would require a finite time in practice; moreover, a very rapid change in the scattering length is not possible since the low momentum approximation for the $T$-matrix scattering potential is no longer valid \cite{Barcelo2003,Fischer2004}. In spite of this, the sudden transition still provides a useful prediction for comparison with the results of the classical field simulations.

%+++++++++++++++++++++++++++++++++++++++++++++++++++++++++++++++++
\subsection{$\tanh$ expansion $($$2+1$ dimensions$)$}\label{sect:tanhacoustic}
%+++++++++++++++++++++++++++++++++++++++++++++++++++++++++++++++++
One non-trivial form of the metric tensor for which the particle production can be calculated analytically is the case of a $\tanh$ function expansion, with asymptotically flat in and out regions. This was first considered by Bernard and Duncan \cite{Bernard1977}, and Birrell and Davies \cite{Birrell1982} for a massive scalar field, and then by Barcel\'{o} \etal~\cite{Barcelo2003} for a massless scalar field. Similarly to \cite{Barcelo2003} we consider the case of $\tanh$ expansion, but for 2+1 dimensions.

In particular (\ref{fieldequation_k_auxiliary}) can be solved exactly when the auxillary factor $\Lambda$ has the time-dependence
\begin{eqnarray}\label{eqn:tanh_omega}
\Lambda^3(\taux) = \frac{\Lambda_i^3 + \Lambda_f^3}{2} + \frac{\Lambda_f^3 - \Lambda_i^3}{2} \tanh \left(\frac{\taux}{\taux_s}\right),
\end{eqnarray}
for some time constant $\taux_s$ that determines the rate of expansion.
Noting that
\begin{eqnarray}
\Lambda^3 = \Omega_{0}^2 \; \frac{1}{b(\taux)},
\end{eqnarray}
the scaling function with respect to auxillary time is
\begin{eqnarray}\label{btaux}
b(\taux) & = & \Omega_{0}^2 \left[\frac{\Lambda_0^3 + \Lambda_f^3}{2} + \frac{\Lambda_f^3 - \Lambda_0^3}{2} \tanh \left(\frac{\taux}{\taux_s}\right)\right]^{-1} \nonumber \\
& = & 2 \left[ 1 + X + (X - 1) \tanh \left(\frac{\taux}{\taux_s}\right)\right]^{-1},
\end{eqnarray}
for an expansion $X$; we have implicitly assumed $b(\taux_i) = 1$.
We can also write
\begin{eqnarray}\label{ttaux}
\!\!\!\!t\!&=&\!\frac{\Omega_{0}}{c_0} \int \frac{1}{b(\taux)} d\taux \nonumber \\
\!&=&\!\frac{\Omega_{0}}{2 c_0}\!\left\{ (1\!+\!X) \taux + (X\!-\!1) \taux_s \log\!\left[\cosh \left(\frac{\taux}{\taux_s}\right)\right] \right\}.
\end{eqnarray}
This is not easy to invert but $b(\taux)$ and $t(\taux)$ define a parametric curve for $b(t)$ --- this relation is required for implementing a $\tanh$ expansion in laboratory time simulations. With the conformal factor given by (\ref{eqn:tanh_omega}) it is possible to calculate the particle production in each mode exactly. Using the result in \cite{Barcelo2003} we get
\begin{eqnarray}\label{Nk_tanh}
N_{\mathbf{k}}^\outss = \frac{\sinh^2\left[\pi k \taux_s \Omega_0 (\sqrt{X} - 1)/2\right]}{\sinh\left[\pi k \taux_s \Omega_0\right] \, \sinh\left[\pi k \taux_s \Omega_0 \sqrt{X}\right]}.
\end{eqnarray}
Note, in the limit $\taux_s \rightarrow 0$ this reduces to the sudden transition result (\ref{Nout_sudden}) as expected.

%+++++++++++++++++++++++++++++++++++++++++++++++++++++++++++++++++
\subsection{de Sitter universe $($$2+1$ dimensions$)$}\label{sect:desitteracoustic}
%+++++++++++++++++++++++++++++++++++++++++++++++++++++++++++++++++
The case of the de Sitter universe is particularly relevant in cosmology. The inflationary model of the early universe is thought to include a de Sitter phase of rapid expansion which ultimately accounts for the inhomogeneities observed in the present universe \cite{Guth1981}. The de Sitter spacetime is a solution to the Einstein's field equations with a positive cosmological constant, and has a high degree of symmetry. It has been shown that an observer moving in a time-like geodesic will measure a thermal spectrum --- this particle production from cosmological horizons being related to the Hawking and Unruh effect (for event and particle horizons respectively).
% being a manifestation of the Hawking effect. 
This result was first derived by Gibbons and Hawking using the path integral formalism \cite{Gibbons1977} and has been subsequently verified by applying the method of Bogoliubov mode mixing \cite{Lapedes1978}.

%+++++++++++++++++++++++++++++++++++++++++++++++++++++++++++++++++
\subsubsection{Scaling function}
%+++++++++++++++++++++++++++++++++++++++++++++++++++++++++++++++++
To map the FRW analogue model to a de Sitter spacetime, the scaling function for the scattering length in laboratory time (which is equivalent to proper time for two dimensions) is of the form
\begin{equation}\label{bdesitter}
b(t) = e^{- t/t_s}
\end{equation}
with the scaling unit $t_s$ that determines the \emph{rate} of expansion. In this case the Hubble parameter (\ref{hubble}) is given by $H = 1/2 t_s$. 

We further consider the transformation to so-called \emph{conformal} time (denoted $\eta$) by $d \eta = \sqrt{b(t)} \, d t$. In this case we get
\begin{eqnarray}
\eta = -2 t_s e^{-t/2 t_s}, \hspace{1cm} t \geq 0.
\end{eqnarray}
The following limits are evident: (i) $\eta = -2 \, t_s$ for $t = 0$; and (ii) $\eta \rightarrow 0$ as $t \rightarrow + \infty$. We also have:
\begin{eqnarray}\label{ds_scaling2}
b(\eta) = \left( \frac{\eta}{2 t_s} \right)^2, \hspace{1cm} -2 t_s \leq \eta \leq 0.
\end{eqnarray}

\begin{figure}[!t]
    \begin{center}
		\includegraphics[width=0.85\columnwidth]{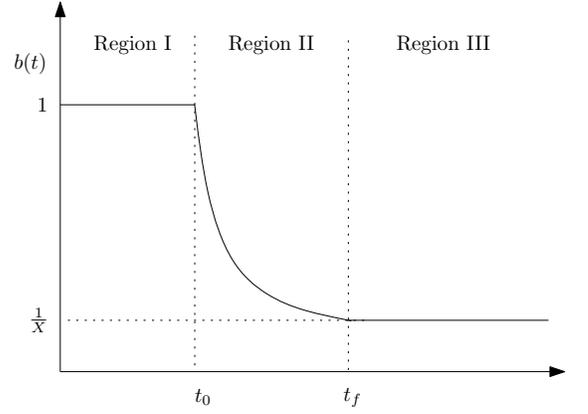}		
		\caption{Schematic of de Sitter expansion for the FRW analogue model; for region I ($t < t_0$), there is no expansion and the mode solutions are those of a Minkowski spacetime with $U = U_0$; for region II ($t_0 < t < t_f$), there is a de Sitter type expansion and the mode solutions are non-trivial; finally for region III ($t > t_f$) the expansion is turned off and the mode solutions are those of a Minkowski spacetime with $U = U_0/X$.}
		\label{desitter_diagram}
		\end{center}
\end{figure}

%~~~~~~~~~~~~~~~~~~~~~~~~~~~~~~~~~~~~~~~~~~~~~~~~~~~~
\subsubsection{Mode solutions}
%~~~~~~~~~~~~~~~~~~~~~~~~~~~~~~~~~~~~~~~~~~~~~~~~~~~~
The field equation (\ref{fieldequation_acoustic}) then yields the second-order differential equation for the mode functions
\begin{eqnarray}
\frac{\partial^2 \chik}{\partial \eta^2} - \frac{1}{\eta} \frac{\partial \chik}{\partial \eta} + c_0^2 k^2 \chik = 0.
\end{eqnarray}
This is a Bessel equation and the solution \cite[Eq. 9.1.52]{abramowitzstegun} is given in terms of Bessel functions of the first and second kind  
%----------------------------------
\footnote{With no loss of generality (as the Bessel equation is even in $\eta$), the arguments of $J_1$ and $Y_1$ are taken to be negative with respect to $\eta$ but positive overall in the region of interest -- this facilitates computation as the Bessel functions are then real quantities. 
}
%----------------------------------
\begin{eqnarray}\label{chikds}
\chik = A_k \, \eta \, J_1(-\omega_k(0) \eta) + B_k \, \eta \, Y_1(-\omega_k(0) \eta)
\end{eqnarray}
for some undetermined constants $A_k$ and $B_k$; we have defined the frequency $\omega_k(0) = c_0 k$ at $t = 0$. 
Using (\ref{nkacoustic}) and \cite[Eq. 9.1.27]{abramowitzstegun} it can be shown that the density fluctuation mode function is given by 
\begin{eqnarray}\label{nkds}
n_k = -2 t_s \frac{\hbar \omega_k(0)}{U_0} \left[ A_k J_0(-\omega_k \eta) + B_k Y_0(-\omega_k \eta) \right].
\end{eqnarray}

With no loss of generality we choose $b(t) = 1$ for $t \leq 0$ --- \emph{i.e.}, for $\eta \leq -2 t_s$. Thus $t = 0$ (or $\eta = -2 t_s$) corresponds to the end of the \emph{in} region, whereas $t \geq t_f$ corresponds to the \emph{out} region (see Fig.~\ref{desitter_diagram}). To determine the particle production it will be necessary to calculate the Bogoliubov coefficients $\alpha_k$ and $\beta_k$ between these two regions.

Using \cite[equation 9.1.16]{abramowitzstegun} it can be shown by matching the Minkowski and de Sitter modes at $t = 0$ that
\begin{eqnarray}
A_k = \frac{\pi}{2} \sqrt{\frac{U_0 \omega_k(0)}{2 \hbar}} \left[ i Y_1(2 t_s \omega_k(0))\!-\!Y_0(2 t_s \omega_k(0))\right],
\end{eqnarray}
\begin{eqnarray}
B_k = - \frac{\pi}{2} \sqrt{\frac{U_0 \omega_k(0)}{2 \hbar}} \left[ i J_1(2 t_s \omega_k(0))\!-\!J_0(2 t_s \omega_k(0))\right]. \nonumber \\
\end{eqnarray}
It can easily be verified using \cite[equation 9.1.16]{abramowitzstegun} that these mode solutions satisfy the normalization condition (\ref{wronskiancommutator}). 

%~~~~~~~~~~~~~~~~~~~~~~~~~~~~~~~~~~~~~~~~~~~~~~~~~~~~
\subsubsection{Particle production}
%~~~~~~~~~~~~~~~~~~~~~~~~~~~~~~~~~~~~~~~~~~~~~~~~~~~~
The lack of a (globally) time-like Killing vector for the de Sitter universe means it is not possible to unambiguously define a Fock vacuum for all times. Additionally, the particle number of each mode after some expansion is observer dependent. However, for our analogue model, we circumvent this complication by associating an instantaneous Minkowski vacuum at each point in time --- that is, we project into the quasiparticle basis that diagonalizes the many body Hamiltonian to second order. This prescription we now follow has been outlined in Sec.~\ref{sect:bogproduction}. Figure \ref{desitter_diagram} shows the scaling factor for the relevant temporal regions.

For convenience we define 
\begin{equation}
R_k(t) \equiv - \omega_k(0) \eta = 2 t_s \omega_k(0) e^{-t/2ts} 
\end{equation}
as well as 
\begin{equation}
\lambda_0 = \sqrt{2 U(0) n_0 / \hbar \omega_k(0)}.
\end{equation} 
Using the mode solutions (\ref{chikds}) and (\ref{nkds}) we then have
\begin{widetext}
\begin{align}\label{nkds2}
\frac{1}{4 n_0} |n_k^{dS}(t)|^2 = \frac{1}{\lambda_0^2} \frac{(\pi t_s \omega_k(0))^2}{4} 
\bigg\{ & \left[Y_1^2(R_k(0)) + Y_0^2(R_k(0))\right] J_0^2(R_k(t)) + \left[J_1^2(R_k(0)) + J_0^2(R_k(0))\right] Y_0^2(R_k(t)) \nonumber \\
&- 2 \left[J_1(R_k(0)) Y_1(R_k(0)) + J_0(R_k(0)) Y_0(R_k(0))\right] J_0(R_k(t)) Y_0(R_k(t)) \bigg\},
\end{align}
and
\begin{align}\label{chikds2}
n_0 |\chi_k^{dS}(t)|^2 = \lambda_0^2 \frac{(\pi t_s \omega_k(0))^2}{4 X} 
\bigg\{ & \left[Y_1^2(R_k(0)) + Y_0^2(R_k(0))\right] J_1^2(R_k(t)) + \left[J_1^2(R_k(0)) + J_0^2(R_k(0))\right] Y_1^2(R_k(t)) \nonumber \\
&- 2 \left[J_1(R_k(0)) Y_1(R_k(0)) + J_0(R_k(0)) Y_0(R_k(0))\right] J_1(R_k(t)) Y_1(R_k(t)) \bigg\}.
\end{align}
\end{widetext}
We can therefore use (\ref{Nkbog}) to calculate the particle production explicitly -- for example, Fig.~\ref{fig:prediction_dsa} shows the particle production for a fixed expansion $X = 2000$ with several different rates of expansion $t_s$.

%~~~~~~~~~~~~~~~~~~~~~~~~~~~~~~~~~~~~~~~~~~~~~~~~~~~~
\subsubsection{Limits}
%~~~~~~~~~~~~~~~~~~~~~~~~~~~~~~~~~~~~~~~~~~~~~~~~~~~~
The particle production from the de Sitter expansion interpolates between two opposite limits: a sudden expansion for $t_s \rightarrow 0$, and an analytically tractable asymptotic limit for $1 \ll 2 t_s \omega_k(0) \ll X^{1/2}$.

% --- non-optimised version --- 
\begin{figure}[!t]
    \begin{center}
	\includegraphics[width=0.85\columnwidth]{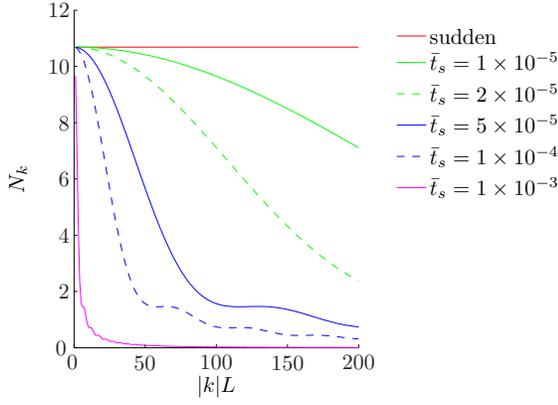}
		%}
		\caption{Particle production for a de Sitter spacetime in the FRW analogue model with $X = 2000$ and a range of expansion rates $\bar{t}_s$ as shown. The dimensionless scaling unit is $\bar{t}_s = t_s \, \hbar/(m L^2)$ and the nonlinearity is $U_0 N_0 = 10^5 \hbar^2/(m L^2)$.}
		\label{fig:prediction_dsa}
		\end{center}
\end{figure}

%-----------------------------------------------------
\begin{itemize}
%-----------------------------------------------------
\item[] {\bf Sudden expansion:}
In this case we have the Minkowski mode functions:
\begin{eqnarray}
\frac{1}{4 n_0} |n_k^{dS}(t)|^2 \rightarrow \frac{1}{4 \lambda_0^2}, \\
n_0 |\chi_k^{dS}(t)|^2  \rightarrow \frac{1}{4} \lambda_0^2,
\end{eqnarray}
and the particle production from (\ref{Nkbog}) reduces to the expression (\ref{sudden_hydro}) we found previously for a sudden expansion
\begin{eqnarray}
N_k \rightarrow \frac{1}{4} \left( X^{1/4} - X^{-1/4} \right)^2.
\end{eqnarray}
This behaviour can be clearly seen in Fig.~\ref{fig:prediction_dsa} where the particle number approaches the sudden result for faster expansion rates (\emph{i.e.}, smaller values of $t_s$).
%-----------------------------------------------------

\item[] {\bf Asymptotic expansion:}
Consider the condition
\begin{equation}\label{dsAsymptCondition}
1 \ll 2 t_s \omega_k(0) \ll X^{1/2},
\end{equation}
which places a constraint on the input frequencies
\begin{eqnarray}
H \ll \omega_k(0) \ll H X^{1/2}.
\end{eqnarray}
In view of the fact that $\omega_k(t) = \omega_k(0)/X^{1/2}$ we can deduce
\begin{eqnarray}\label{dsAsymptCondition2}
X^{-1/2} \ll 2 t_s \omega_k(t) \ll 1,
\end{eqnarray}
and so rewrite the initial constraint (\ref{dsAsymptCondition}) and its consequence (\ref{dsAsymptCondition2}) as
\begin{eqnarray}
1 \ll R_k(0) \ll X^{1/2}, \\
X^{-1/2} \ll R_k(t) \ll 1.
\end{eqnarray}
In terms of the expansion $X$ we can rewrite (\ref{Nkbog}) as
\begin{eqnarray}
N_k = \frac{\lambda_0^2}{\sqrt{X}} \frac{1}{4 n_0} |n_k^{\rm exp}|^2 + \frac{\sqrt{X}}{\lambda_0^2} n_0 |\chi_k^{\rm exp}|^2 - \half
\end{eqnarray}
Then using (\ref{nkds2}) and (\ref{chikds2}) we have 
\begin{widetext}
\begin{align}
N_k = & \left( \frac{\pi}{2} \right)^2 \frac{(2 t_s \omega_k(0))^2}{4 \sqrt{X}} \times \nonumber \\
& \bigg\{ \left[Y_1^2(R_k(0)) + Y_0^2(R_k(0))\right] \left[J_0^2(R_k(t)) + J_1^2(R_k(t))\right]
+ \left[J_1^2(R_k(0)) + J_0^2(R_k(0))\right] \left[Y_0^2(R_k(t)) + Y_1^2(R_k(t))\right] \nonumber \\
& - 2 \left[J_1(R_k(0)) Y_1(R_k(0)) + J_0(R_k(0)) Y_0(R_k(0))\right] \left[J_0(R_k(t)) Y_0(R_k(t)) + J_1(R_k(t)) Y_1(R_k(t)) \right] \bigg\} - \frac{1}{2}
\end{align}
\end{widetext}
Now using the limits for Bessel functions given in Abramowitz and Stegun \cite{abramowitzstegun}, as well as some basic trigonometric identities, we deduce
\begin{eqnarray}
N_k \rightarrow \frac{X^{1/2}}{4 \pi \; t_s \; \omega_k(0)} = {1\over4\pi t_s \omega_{k}}.
\end{eqnarray}
Further noting that the low $k$ (and/or high temperature) limit of the Bose-Einstein distribution is 
\begin{equation}
N_k \approx k_B T/\hbar \omega_k
\end{equation}
leads to the temperature
\begin{eqnarray}\label{desittertemp2}
k_B T = \frac{1}{4 \pi t_s},
\end{eqnarray}
in units of $\hbar = c(0) = 1$. That is, 
the particle production approaches the usual cosmological result of a thermal spectrum \cite{Birrell1982} for an sufficiently large expansion that proceeds sufficiently slowly. 

\end{itemize}
%--------------------------------------------------------

%%%%%%%%%%%%%%%%%%%%%%%%%%%%%%%%%%%%%%%%%%%%%%%%%%%
%
\section{Quasiparticle production beyond the acoustic approximation}\label{sect:nonlindispresults}
%
%%%%%%%%%%%%%%%%%%%%%%%%%%%%%%%%%%%%%%%%%%%%%%%%%%%
In this section we consider quasiparticle production where the acoustic approximation is not enforced --- we first introduced this more general case in section \ref{lorentzbreak}. In general the usual calculations become unmanageable in this regime, and so, we present only two expansion scenarios here: We first calculate quasiparticle production for a sudden transition in Sec.~\ref{sect:suddenfull}, for which an exact solution can be easily found. In Sec.~\ref{sect:cyclicacoustic} we additionally consider the cyclic universe model, for which we are able to make some general statements with regards to quasiparticle production.

%+++++++++++++++++++++++++++++++++++++++++++++++++++++++++++++++++
\subsection{Sudden Expansion}\label{sect:suddenfull}
%+++++++++++++++++++++++++++++++++++++++++++++++++++++++++++++++++
Particle production for a sudden expansion can be calculated using the Bogoliubov theory from Sec.~\ref{sect:bogconnection}. The scaling function is given by (\ref{bsudden}) with the substitution for laboratory time $\taux \rightarrow t$. The excitation of each mode is attributed to the quantum depletion corresponding to the initial vacuum state with nonlinearity $U_0$, projected into the Bogoliubov basis corresponding to the final nonlinearity $U_0/X$.
We can thus calculate the number of Bogoliubov quasiparticles directly using (\ref{bog_transform}), assuming the initial state is the Bogoliubov vacuum defined by $\hat{b}^{\inss}_{\mathbf{k}} | 0 \rangle = 0$. Using (\ref{Nkbog}) the result  is
\begin{eqnarray}\label{Nout_sudden3}
N_{\mathbf{k}}^{\outss} & = & (u_{\mathbf{k}}^{\outss} v_{\mathbf{k}}^{\inss}
- v_{\mathbf{k}}^{\outss} u_{\mathbf{k}}^{\inss})^2 \nonumber \\
& = & \frac{(A_{\mathbf{k}}^{\inss} - A_{\mathbf{k}}^{\outss})^2}{(1 - A_{\mathbf{k}}^{\inss \, 2}) (1 - A_{\mathbf{k}}^{\outss \, 2})},
\end{eqnarray}
where $A_{\mathbf{k}}^{\inss}$ and $A_{\mathbf{k}}^{\outss}$ is given by (\ref{bogAk}) with $U = U_0$ and $U = U_0/X$ respectively. The particle production from (\ref{Nout_sudden3}) is suppressed for modes of large momenta, as can be seen from (\ref{bogAk}) by the observation that $A_{\mathbf{k}}^{\inss/\outss} \rightarrow 0$ as $|\mathbf{k}| \rightarrow \infty$; --- this is consistent with the preceding discussion of Sec.~\ref{lorentzbreak}.

This calculation can also be repeated using the usual methods of quantum field theory in curved spacetime as outlined in Sec.~\ref{sect:quantisation} --- similarly to the previous calculation for the sudden transition in the acoustic approximation outlined in Sec.~\ref{sect:suddenacoustic} --- but instead by using the field equation (\ref{fieldequation_nonlin}), which includes nonlinear dispersion.  The result is \cite{Weinfurtner:2007ab}
\begin{eqnarray}\label{Nout_sudden2}
N_{\mathbf{k}}^\outss = |\beta_{\mathbf{k}}|^2 = \quarter \left ( 
\sqrt{\frac{\omega_{\mathbf{k}}^{\inss}}{\omega_{\mathbf{k}}^{\outss}}}
-  \sqrt{\frac{\omega_{\mathbf{k}}^{\outss}}{\omega_{\mathbf{k}}^{\inss}}}
\right )^2.
\end{eqnarray}
This takes the same form as the result from the acoustic approximation given by (\ref{Nout_sudden}). However, here $\omega_{\mathbf{k}}^{\inss}$ and $\omega_{\mathbf{k}}^{\outss}$ are the Bogoliubov mode frequencies given by (\ref{bogdispersion}), which includes nonlinear dispersion, for the \emph{in} and \emph{out} regions respectively. 

It is straightforward to show that (\ref{Nout_sudden3}) and (\ref{Nout_sudden2}) agree exactly for all $|\mathbf{k}|$.
We further note that for small $|\mathbf{k}|$, the Bogoliubov coefficients are $u_k \approx v_k$ and the Bogoliubov mode expansion coincides with the mode expansion for the quantised phase fluctuations (\ref{mode_expansion}). Therefore in this limit the acoustic result (\ref{sudden_hydro}) agrees with (\ref{Nout_sudden3}).

%+++++++++++++++++++++++++++++++++++++++++++++++++++++++++++++++++
\subsection{Cyclic universe $($$2+1$ dimensions$)$}\label{sect:cyclicacoustic}
%+++++++++++++++++++++++++++++++++++++++++++++++++++++++++++++++++
The final scenario we consider is that of a \emph{cyclic} (or \emph{oscillatory}) universe. While models of a cyclic universe certainly exist in the literature \cite{Steinhardt2002}, they are not as firmly established as inflationary models (such as the de Sitter universe). From a condensed matter point of view, an analogue model of a cyclic universe is interesting because it leads to parametric excitation of the quasiparticle modes, and in particular to parametric resonance \cite{MechanicsLL}. Moreover, by taking $b(t_f) = b(t = 0)$, a cyclic universe is an interesting counter-point to the case of a sudden transition: in a sudden transition the field does not evolve and any particle production is entirely attributed to a sudden change in the effective spacetime (\emph{i.e.}, we project into a new quasiparticle basis that depends on the final nonlinearity $U = U_0/X$); however, in a cyclic universe model, the initial and final effective spacetimes are the same (so long as condensate does not evolve too far from the initial ground state) and particle production occurs due to parametric excitation only. 

A suitable function can be expressed as 
\begin{eqnarray}\label{eqn:scalecyclic}
b(t) = \frac{1}{X} + \frac{1}{2} \left( 1 - \frac{1}{X}  \right) \left[\cos \left( \frac{2\pi m t}{t_f} \right) + 1 \right]
\end{eqnarray}
with $b(0) = 1$ and where we have defined $m$ as the number of cycles of the oscillation (we note that $t_s = t_f/m$ is the period of each oscillation), and $X$ is now defined as the amplitude of the oscillation (rather than the final expansion).

%~~~~~~~~~~~~~~~~~~~~~~~~~~~~~~~~~~~~~~~~~~~~~~~~~~~~
\subsubsection{Parametric resonance}
%~~~~~~~~~~~~~~~~~~~~~~~~~~~~~~~~~~~~~~~~~~~~~~~~~~~~
The phenomena of parametric resonance has been previously investigated in Bose condensed systems \cite{GarciaRipoll1999, Modugno2006}, and in the context of an expanding universe model \cite{Zlatev1998,Bassett1999}.
The condition for parametric resonance occurs close to $\omega(k) = \Omega/2$ where $\Omega = 2\pi m/t_f$ is the driving frequency of some external parameter \cite{MechanicsLL}, in this case the nonlinearity. Using the Bogoliubov excitation spectrum $\epsilon(k) = \hbar \omega(k)$ from (\ref{bogdispersion}) then gives a simple estimate of the peak wave-vector for resonance:
\begin{eqnarray}\label{cyclicresonance}
k_{\textrm{res}}\!=\!\left(\frac{2m}{\hbar^2}\right)^{1/2}\!\!\left\{ [(U_{\textrm{ave}} N_0)^2\!+\!(\hbar \pi/t_s)^2]^{1/2} \!-\!U_{\textrm{ave}} N_0 \right\}^{1/2} \nonumber \\
\end{eqnarray}
where $U_{\textrm{ave}} = \half (1 + 1/X) U_0$ is the average nonlinearity during the evolution. However, it is worth noting that this result requires that $\Delta = 1 - 1/X$ from (\ref{eqn:scalecyclic}) is a perturbative parameter with $\Delta \ll 1$; we therefore henceforth refer to this as the \emph{perturbative resonance condition}.

The extension of the analysis to a strongly driven system ($\Delta \lesssim 1$) has been considered in \cite{GarciaRipoll1999}. The general result is that the mode frequency region where parametric resonance occurs broadens as $\Delta$ increases; that is, for the parametric resonance peak broadens in momentum space with larger $X$. 

%%%%%%%%%%%%%%%%%%%%%%%%%%%%%%%%%%%%%%%%%%%%%%%%%%%
%
\section{The classical field method}\label{chap:cfm}
%
%%%%%%%%%%%%%%%%%%%%%%%%%%%%%%%%%%%%%%%%%%%%%%%%%%%

Classical field methods are a powerful tool for approximating the dynamics of quantum systems. Their application to BECs include the closely related methodologies of the finite temperature Gross--Pitaevskii equation \cite{Davis2002,blakie2004}, the truncated Wigner approximation \cite{Steel1998,Sinatra2000,Sinatra2001,Sinatra2002,norrie2005,norrie2006} and the positive-P method \cite{Steel1998,Drummond1999}.

Neglecting quantum and thermal fluctuations, the condensate dynamics are determined by the GPE (\ref{eq:gpe}).  This approximation arises from the assumption that the condensate is highly occupied so that $ N_0 = \int d\x \langle \hat{\psi^\dagger}(\x) \hat{\psi}(\x) \rangle \gg 1$. 
In the classical field approximation this description is extended by also including the non-condensate modes from a low energy subspace of the system; these modes are to be considered classical in that they are highly populated --- this is akin to the Bogoliubov approximation where the commutators can be neglected. We thus proceed by expanding the field operator in some basis
\begin{eqnarray}\label{psiexpansion3}
\hat{\psi}(\x, t) = \sum_{\mathbf{k}} \phi_{\mathbf{k}}(\x) \; \hat{a}_{\mathbf{k}} (t),
\end{eqnarray}
and replacing this in (\ref{heisenbergfield}) by the classical field
\begin{eqnarray}\label{classicalfield}
\psi(\x, t) = \sum_{{\mathbf{k}} \in C} \phi_{\mathbf{k}}(\x) \; \alpha_{\mathbf{k}} (t),
\end{eqnarray}
where $\hat{a}_{\mathbf{k}} \rightarrow \alpha_{\mathbf{k}}$ for those modes where $N_{\mathbf{k}} = \langle \hat{a}_{\mathbf{k}}^{\dagger} \hat{a}_{\mathbf{k}} \rangle \gg 1$ is satisfied; we denote these modes in the low-energy subspace by $k \in C$. For a homogeneous Bose gas in a box with periodic boundary conditions the modes of the system (the eigenstates that diagonalize the single-particle Hamiltonian) are plane wave states
\begin{eqnarray}
\phi_{\mathbf{k}}(\x) = \frac{1}{\sqrt{V}} e^{i \mathbf{k} \cdot \x}.
\end{eqnarray}
The low-energy subspace $C$ is then determined by a momentum cut-off, below which all modes are retained in the classical field. This is formalised by the use of a projector which is defined by its action on some function $f(\x)$ as
\begin{eqnarray}
\hat{P} \{ f(\x) \} = \sum_{\mathbf{k} \in C} \phi_{\mathbf{k}}(\x) \int d\x^{\prime} \,
\phi_{\mathbf{k}}^*(\x^{\prime}) f(\x^{\prime}).
\end{eqnarray}

%+++++++++++++++++++++++++++++++++++++++++++++++++++++++++++++++++
\subsection{Projected Gross-Pitaevskii equation}
%+++++++++++++++++++++++++++++++++++++++++++++++++++++++++++++++++
Neglecting all modes orthogonal to $C$ the projected Gross-Pitaevskii equation (PGPE) is given by
\begin{eqnarray}\label{pgpe}
i \hbar \frac{\partial \psi(\x)}{\partial t} = -\frac{\hbar^2}{2 m} \nabla^2\ \psi(\x) + U_0 \hat{P} \{|\psi(\x)|^2 \psi(\x)\}.
\end{eqnarray}
For consistency with our FRW analogue model we have not included an external potential here.
The projector is required for the following reasons:
\begin{enumerate}
\item[(i)] The classical field approximation naturally divides the system into a coherent region, which is described by the propagation of a classical field, and an incoherent region which is neglected in the present formalism. In equilibrium, the system is then described by a microcanonical ensemble since particle numbers are conserved by the Hamiltonian (\ref{secondquantH}).
\item[(ii)] While the finite size of the spatial grid inherently defines a momentum cut-off, the split operator Fourier methods used to propagate the classical field can introduce aliasing if a projector is not applied explicitly.
\item[(iii)] The application of a projector is consistent with using a contact potential to describe the two-body interactions; this description leads to ultraviolet divergences at large momenta and so a cut-off is required.
\end{enumerate}

%+++++++++++++++++++++++++++++++++++++++++++++++++++++++++++++++++
\subsection{The truncated Wigner approximation}\label{sect:twa}
%+++++++++++++++++++++++++++++++++++++++++++++++++++++++++++++++++
A formal framework for the ideas outlined above is provided by the truncated Wigner approximation (TWA). We briefly outline the method but the reader is referred to \cite{Steel1998,Sinatra2000,Sinatra2001,Sinatra2002,norrie2005,norrie2006} for further details. 

The TWA is a phase space method originating from the representation of the density operator in terms of the Wigner function, which is familiar from quantum optics \cite{Gardiner2004}. The master equation for the multimode density operator can be formally mapped to a third-order differential equation for the Wigner function by the application of operator correspondences. The approximation involved in the TWA is to neglect the third-order derivative terms, which become small for highly occupied modes. The resulting Fokker--Planck type equation has no diffusion term and is equivalent to evolving a classical field of the form (\ref{classicalfield}) with the GPE, however with two crucial modifications:
\begin{enumerate}
\item  Quantum vacuum fluctuations are included in the initial state by adding classical noise sampled from the Wigner distribution; the form of this noise depends on the Wigner function for the ground state of the system. For a BEC at $T = 0$ the initial amplitude of each mode is a random Gaussian variable that is distributed according to the Wigner function for a coherent state.
\item The moments of the the Wigner function give the expectation values for symmetrically ordered operators. In practice calculating the expectation value of an observable $O$ requires an ensemble average over many trajectories in phase space. We denote such an expectation value by $\langle O \rangle_W$. 
\end{enumerate}

The initial condition is given by superposition of the ground state and noise sampled from the Wigner distribution.
\begin{eqnarray}
\psi(\x, t = 0) = \psi_0(\x) + \delta\psi(\x).
\end{eqnarray}
We can expand the noise term via a Fourier transform as
\begin{eqnarray}
\eta(\mathbf{r}) = \frac{1}{\sqrt{V}} \sum_{\mathbf{k} \neq 0} e^{i \mathbf{k} \cdot \mathbf{r}} \eta_{\mathbf{k}}.
\end{eqnarray}

Within the truncated Wigner approximation the initial vacuum state
is prepared by specifying noise on each of the Bogoliubov modes:
\begin{eqnarray}
\delta\psi(\x) = \sum_{k \neq 0, k \in C} \left ( U_{\mathbf{k}}(\x) \beta_{\mathbf{k}} + 
V^*_{\mathbf{k}}(\x) \beta^*_{\mathbf{k}} \right ),
\end{eqnarray}
where $U_{\mathbf{k}}(\x)$ and $V_{\mathbf{k}}(\x)$ are the plane wave modes with amplitudes $u_{\mathbf{k}}$ and $v_{\mathbf{k}}$ respectively (as defined in Sec.~\ref{bogtheory}). The $\beta_{\mathbf{k}}$ are complex random variables that obey the Gaussian statistics \cite{Gardiner2002,Steel1998}:
\begin{eqnarray}
\langle \beta_{\mathbf{p}} \beta_{\mathbf{q}} \rangle &=& \langle \beta^*_{\mathbf{p}} \beta^*_{\mathbf{q}} \rangle = 0, \label{eq:twaguassian1} \\
\langle \beta_{\mathbf{p}}^* \beta_{\mathbf{q}} \rangle &=& \frac{1}{2} \delta_{p, q} \label{eq:twaguassian2}.
\end{eqnarray}

The initial state is thus constructed by populating the Bogoliubov modes with half a particle per mode according to the TWA prescription, for the initial nonlinearity $U_0$; for our cosmological model this corresponds to the instantaneous vacuum state (Minkowski vacuum) in laboratory time at $t = 0$.

The Wigner and quantum expectation values for the population of the $\mathbf{k}$ mode in the Bogoliubov quasiparticle basis are related by
\begin{eqnarray}\label{twaqpnumber}
\langle \beta_{\mathbf{k}}^* \beta_{\mathbf{k}} \rangle_{W} = \langle \{ \hat{b}_{\mathbf{k}}^\dagger \hat{b}_{\mathbf{k}} \} \rangle = \langle \hat{N}_{\mathbf{k}} \rangle + \frac{1}{2},
\end{eqnarray}
where the braces require that the symmetrised operator should be taken.
The vacuum state therefore corresponds to half a particle per mode in the classical field. 

%+++++++++++++++++++++++++++++++++++++++++++++++++++++++++++++++++
\subsection{Validity of the TWA}\label{twavalidity}
%+++++++++++++++++++++++++++++++++++++++++++++++++++++++++++++++++
In the present application of the TWA, only the $\mathbf{k} = 0$ condensate mode is macroscopically occupied, the other modes being initially unpopulated (\emph{i.e.}, the quantum expectation value is $\langle \hat{N}_{\mathbf{k}} \rangle = 0$). The requirement that $N_{\mathbf{k}} \gg 1$ for each mode in the classical field is then violated. However a more detailed treatment of the validity of the TWA leads to the criterion that $N \gg M$ for a system of $N$ particles and $M$ modes \cite{Sinatra2002}. This criterion has been made explicit by Norrie \etal~\cite{norrie2006} as the requirement that the particle density exceeds the commutator for the restricted field operator . It has been shown that for a homogeneous system these two criteria coincide \cite{norriephd}.  Therefore the TWA can still be applied when most of the modes are unoccupied as long as the average particle density is sufficiently large.

Moreover the above choice for the initial state can lead to heating as is evident by a transient thermalization of the system \cite{Steel1998, Sinatra2000}. In this case the classical field dynamics deviate from the Bogoliubov theory valid for a weakly interacting gas; the system evolves to thermal equilibrium via the nonlinear interactions between Bogoliubov modes. This effect can be suppressed by evolving the classical field only for short times and by choosing a regime where system is weakly interacting (\emph{i.e.}, $U$ small). This is an important consideration for our simulations where any thermalization could obscure the effect of particle production.

\subsection{Quasiparticle number}\label{sect:twaqpnumber}

Following the discussion on Bogoliubov theory in Sec.~\ref{sect:bogconnection}, the classical field for the homogeneous system can be expressed as
\begin{eqnarray}
\psi(\mathbf{x}, t) &=& e^{-i \mu t/\hbar}
\left[\psi_0(\mathbf{x}) + \delta \psi(\mathbf{x}, t) \right], \nonumber \\
\delta \psi(\mathbf{x}, t) &=& 
\frac{e^{i \mathbf{k} \cdot \mathbf{x}}}{\sqrt{V}} \sum_k \left(u_k \beta_{-\mathbf{k}}(t)  + v_k \beta^*_{\mathbf{k}}(t) \right)
\end{eqnarray}
where the time-dependent amplitudes are given from (\ref{bog_transform}) by
\begin{eqnarray}
\beta_{\mathbf{k}}(t) = u_{\mathbf{k}} \alpha_{\mathbf{k}}(t) \; \frac{\alpha_0^*}{|\alpha_0|}
 - v_{\mathbf{k}} {\alpha^*}_{-\mathbf{k}}(t) \; \frac{\alpha_0}{|\alpha_0|},
\end{eqnarray}
and where the condensate phase factor is given by $\alpha_0/|\alpha_0| = e^{-i \mu t/\hbar}$.
Referring to (\ref{twaqpnumber}), the quasiparticle number in the TWA prescription (where the expectation value is implicitly assumed) is given by
\begin{eqnarray}\label{Nkcfm}
N_\mathbf{k}(t) = \langle {\beta^*}_{\mathbf{k}}(t) \, \beta_{\mathbf{k}}(t) \rangle_{W} - \half.
\end{eqnarray}
We use this quantity to calculate the quasiparticle mode populations in our simulations; at each time the mode functions $u_k$ and $v_k$ are determined from (\ref{bogukvk}) using nonlinearity $U(t)$. Hence this result is consistent with (\ref{Nkbog}), as it requires projection into the quasiparticle basis that instantaneously diagonalizes (to second-order in quasiparticle operators) the many body Hamiltonian (\ref{secondquantH}).

%+++++++++++++++++++++++++++++++++++++++++++++++++++++++++++++++++
\subsection{Numerical details}
%+++++++++++++++++++++++++++++++++++++++++++++++++++++++++++++++++

Our ``universe" is specified by a box with dimensions $L_x = \gamma_x L$, $L_y = \gamma_y L$, and $L_z = \gamma_z L$. 
In what follows we assume $\gamma_x = \gamma_y = 1$ and that $\gamma_z$ is strictly less than one as required by the quasi-two-dimensional model. To facilitate numerical computation we introduce the dimensionless parameters
\begin{eqnarray}
\overline{\mathbf{x}} = \frac{\mathbf{x}}{L} \textrm{,} \hspace{0.5cm}
\overline{\psi} = \psi \frac{L^{d/2}}{\sqrt{N_0}} \textrm{,} \hspace{0.5cm}
\overline{t} = t \; \frac{\hbar}{m L^2}.
\end{eqnarray}
With a time-dependent nonlinear interaction, the two-dimensional GPE (\ref{eq:gpe2d}) then takes the dimensionless form
\begin{eqnarray}\label{gpe_dimensionless}
i \frac{\partial \overline{\psi}}{\partial \overline{t}} = \left [-\frac{1}{2} \overline{\nabla}^2 + C_{NL}(\overline{t}) |\overline{\psi}|^2 \right ] \overline{\psi}.
\end{eqnarray}
We have taken $\overline{V}_{\scriptsize\textrm{ext}} = 0$ for the homogeneous system.
The effective nonlinearity is (integrating over the $z$ direction for the quasi-two-dimensional geometry) 
\begin{eqnarray}\label{eqn:dimensionlessCnl}
C_{NL}(\overline{t}) = \frac{U_{\textrm{2D}} \, b(\overline{t}) \, N_0\, m}{\hbar^2} = \frac{4 \pi \, a(\overline{t}) \, N_0}{L_z}.
\end{eqnarray}
Note that the wave function is normalized to unity here. The corresponding dimensionless speed of sound is
\begin{eqnarray}
\overline{c} =  \frac{m\, L}{\hbar} c = \sqrt{C_{NL}}.
\end{eqnarray}
For completeness, we note the Bogoliubov excitation spectrum (\ref{bogdispersion}) in dimensionless units is given by
\begin{eqnarray}\label{bogdispersion_dimless}
\epsilon_{\overline{k}} = \sqrt{\frac{\overline{k}^2}{2} \left(\frac{\overline{k}^2}{2} + 2 C_{NL} \right)}.
\end{eqnarray} 

In dimensionless units the spatial coordinates span the region $-\half \leq \overline{\mathbf{x}} \leq \half$. For convenience we henceforth drop the bar notation (unless otherwise specified).

Equation (\ref{gpe_dimensionless}) is propagated using the 4th order Runge--Kutta algorithm in the interaction picture \cite{caradocdavies}. For the results presented here the time step was chosen so that the total normalization change during each trajectory was $\Delta \textrm{norm} \leq 10^{-9}$ (for our choice of total particle number, this corresponds to a total loss or gain of much less than one particle for the entire field).  

The field is discretized on a grid of $128 \times 128$ points. The projector retains all modes with $|k| \leq 32 \times 2\pi$, an area which includes $M = 3209$ modes in the classical field. 

In practice, the mode populations $N_{\mathbf{k}}(k_x, k_y)$ were resampled on polar coordinates $|k|$ and $\phi$ and then averaged over angle.

%+++++++++++++++++++++++++++++++++++++++++++++++++++++++++++++++++
\subsection{Suitable parameter regime}\label{sect:cfmsuitable}
%+++++++++++++++++++++++++++++++++++++++++++++++++++++++++++++++++
It is appropriate, at this point, to determine a viable set of parameters for the classical field simulations. 
The choice of simulation parameters is constrained by three main factors:
\begin{enumerate}
\item[(i)] The criterion for the validity of the classical field method (\emph{i.e.}, for the TWA).
\item[(ii)] The requirement that all modes of the system are in the phononic regime at the start of the simulation. In  this regime the particle production is significant.
\item[(iii)] A set of parameters that are experimentally relevant.
\end{enumerate}

The criteria for the validity of the TWA has been discussed in Sec.~\ref{twavalidity}. In particular, for our simulations, we are required to choose a condensate population with $N_0 \gg M = 3209$. Additionally, noting that in our simulations the classical field is normalised to unity, the requirement that the system is weakly interacting is satisfied when the nonlinearity $C_{NL}$ is small compared with the condensate population $N_0$. 

We wish to investigate a regime where a significant fraction of the modes are phononic (as determined by $k < 1/\xi$) so that we can compare our results with the analytic calculations in the acoustic approximation. In computational units the phonon to free-particle cross-over is determined by $\overline{k}_c^2/2 = C_{NL}(0)$. That is, we require a large initial nonlinearity.

Recent experimental observations indicate that $^{85}\textrm{Rb}$ condensates have the most widely tunable interactions via a Feshbach resonance. We refer in particular to experimental results from the JILA group \cite{roberts1998,cornish2000}. In their results (see \cite{cornish2000}) a stable condensate of $10^4$ atoms was formed with a variation of the scattering length from zero to $4000 \, a_0$. The associated diluteness factor $n a^3 \sim 10^{-2}$ indicates such a system has significant interactions but can still be considered to be \emph{weakly interacting}.

We employ the parameters from \cite{cornish2000}, but use a larger atom number of $N_0 = 10^7$ while considering the same (peak) number density $N_0/V \approx 10^{12}$ cm$^{-3}$. We also take the maximum possible initial scattering length to be $a = 4000 \, a_0$ at $t = 0$.

With these parameters in mind we calculate the dimensionless parameters required for the classical field simulation. From (\ref{eqn:dimensionlessCnl}) and assuming $L_z \approx \gamma_z V^{1/3}$ we can estimate the (dimensionless) initial nonlinear interaction strength is
\begin{eqnarray}
C_{NL} = \frac{4\pi a}{\gamma_z V^{1/3}} N_0 \approx \frac{1.24 \times 10^5}{\gamma_z}.
\end{eqnarray}
The anisotropy parameter should be taken $\gamma_z < 1$ for the quasi-two-dimensional geometry --- we do not impose a specific value, but note that we are free to choose a value of the scattering length less than $\sim 4000 \, a_0$. 
Therefore to meet all the above requirements we select $C_{NL}(\overline{t} = 0) = 1 \times 10^5$ and $N_0 = 10^7$ for the simulation results presented in this paper. While a stable condensate with this atom number has not yet been achieved experimentally, it gives a diluteness factor less than $n a^3 \sim 10^{-2}$ as found in \cite{cornish2000}, and so should in principle be possible.  

With a large value for $N_0$, we have checked that the thermalization that can occur in the TWA at large nonlinearities is suppressed. The quasiparticle production demonstrated is then due solely to the effects of \emph{expanding} the effective spacetime. Additionally we note that the TWA is valid for short times only --- this allows us to explore systems undergoing rapid expansion and for which there is appreciable particle production.

\begin{figure}[!t]
\begin{center}
\includegraphics[width=0.49\textwidth]{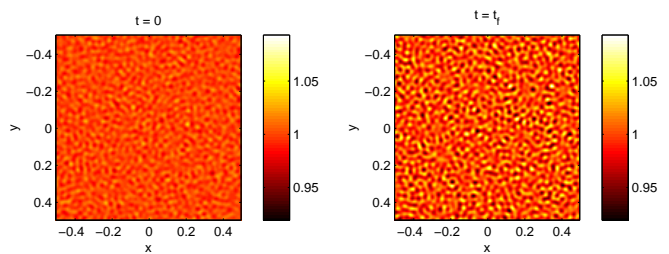}
\caption{de Sitter model: density plot for renormalized wavefunction at beginning ($t = 0$) and at end ($t = t_f$) of expansion. Parameters are $t_s = 1 \times 10^{-5}$, $C_{NL}(t = 0) = 1 \times 10^5$, and $N_0 = 10^7$.}\label{fig:desittersnapshot}
\end{center}
\end{figure}

%%%%%%%%%%%%%%%%%%%%%%%%%%%%%%%%%%%%%%%%%%%%%%%%%%%
%
\section{Expanding Universe simulations}\label{chap:expandingsimulations}
%
%%%%%%%%%%%%%%%%%%%%%%%%%%%%%%%%%%%%%%%%%%%%%%%%%%%
\begin{figure*}[!htb]
    \begin{center}
		\mbox{
		\subfigure[$\; t_s = 1 \times 10^{-5}$]{\includegraphics[width=0.35\textwidth,clip]{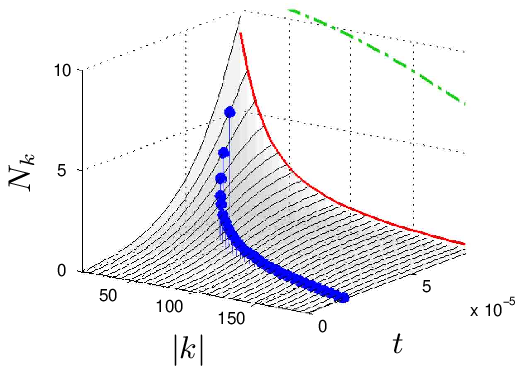}}
		\hspace{5mm}
		\subfigure[$\; t_s = 5 \times 10^{-5}$]{\includegraphics[width=0.35\textwidth,clip]{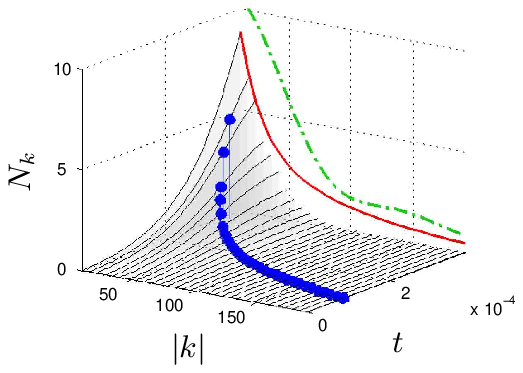}}
		}
		\vspace{0mm}
		\mbox{
		\subfigure[$\; t_s = 1 \times 10^{-4}$]{\includegraphics[width=0.35\textwidth,clip]{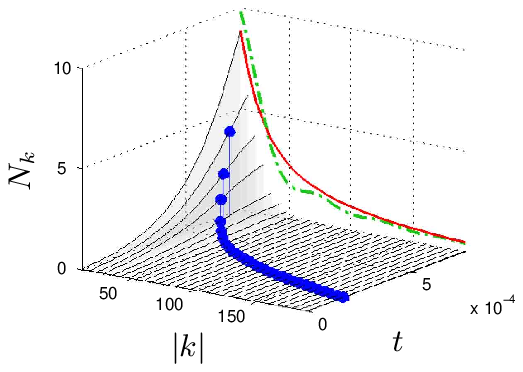}}
		\hspace{5mm}
		\subfigure[$\; t_s = 1 \times 10^{-3}$]{\includegraphics[width=0.35\textwidth,clip]{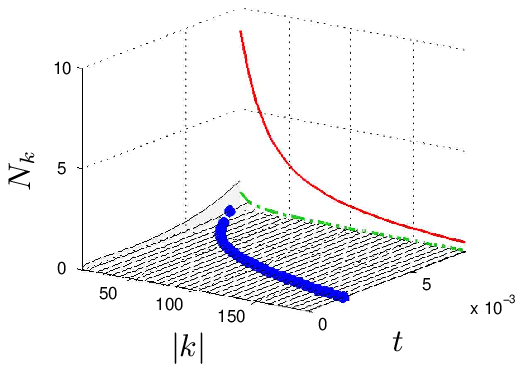}}		}
		\caption{de Sitter expansion: time dependence of Bogoliubov mode populations. Parameters are $C_{NL}(\overline{t} = 0) = 1 \times 10^5$, $N_0 = 10^7$ and $X = 2 \times 10^3$. The blue points on each curve show where each mode crosses over from phonon to free-particle behaviour (as defined by $\overline{k}_c^2/2 = C_{NL}$). The green dashed curve shows the analytic prediction in the acoustic approximation based on the Bogoliubov theory and Hamiltonian diagonalization as outlined in Sec.~\ref{sect:suddenacoustic}. The red solid curve shows the analytic prediction (including quantum pressure) for a sudden transition from (\ref{Nout_sudden3}).}\label{fig:desitterresults}
		\end{center}
\end{figure*}

%% de sitter results - large (for Phys Rev A)
%\begin{figure*}[!htb]
%    \begin{center}
%		\mbox{
%		\subfigure[$\; t_s = 1 \times 10^{-5}$]{\includegraphics[width=0.4\textwidth]{Figures/DeSitter/modes_ds_ts_1e-5.eps}}
%		\hspace{0mm}
%		\subfigure[$\; t_s = 5 \times 10^{-5}$]{\includegraphics[width=0.4\textwidth]{Figures/DeSitter/modes_ds_ts_5e-5.eps}}
%		}
%		\mbox{
%		\subfigure[$\; t_s = 1 \times 10^{-4}$]{\includegraphics[width=0.4\textwidth]{Figures/DeSitter/modes_ds_ts_1e-4.eps}}
%		\hspace{0mm}
%		\subfigure[$\; t_s = 1 \times 10^{-3}$]{\includegraphics[width=0.4\textwidth]{Figures/DeSitter/modes_ds_ts_1e-3.eps}}		}
%		\caption{de Sitter expansion: time dependence of Bogoliubov mode populations. Parameters are $C_{NL}(\overline{t} = 0) = 1 \times 10^5$, $N_0 = 10^7$ and $X = 2 \times 10^3$. The blue points on each curve show where each mode crosses over from phonon to free-particle behaviour (as defined by $\overline{k}_c^2/2 = C_{NL}$). The green dashed curve shows the analytic prediction in the acoustic approximation based on the Bogoliubov theory and Hamiltonian diagonalization as outlined in Sec.~\ref{sect:suddenacoustic}. The red solid curve shows the analytic prediction (including quantum pressure) for a sudden transition from (\ref{Nout_sudden3}).}\label{fig:desitterresults}
%		\end{center}
%\end{figure*}
%%%%%%%%%%%%%%%%%%%%%%%%%%%%%%%%%%%%%%%%%%%%%%%%%%%

We now present the numerical results of classical field simulations based on the TWA for the expansion scenarios outlined in Sec.~\ref{sect:acousticpredictions} --- namely the de Sitter expansion, $\tanh$ and cyclic expansion scenarios, with the sudden transition as a limit of an infinitely fast de Sitter or $\tanh$ expansion. The results are compared to the analytic predictions in the acoustic approximation, and also to the sudden transition prediction that includes the nonlinear dispersion of the modes (see Eq. \ref{Nout_sudden3}).

In the results shown, we have calculated the quasiparticle populations for each mode as a function of time; this was accomplished by projecting from the single particle basis to the Bogoliubov basis using the expression (\ref{Nkcfm}), and using the nonlinearity $C_{NL}(\overline{t})$. Thus the basis for counting quasiparticles corresponds to projecting into the instantaneous Minkowski vacuum at each time. 

\subsection{de Sitter universe}

%
% tanh results
\begin{figure*}[!t]
    \begin{center}
		\mbox{
		\subfigure[$\; \taux_s = 1 \times 10^{-1}$]{\includegraphics[width=0.35\textwidth,clip]{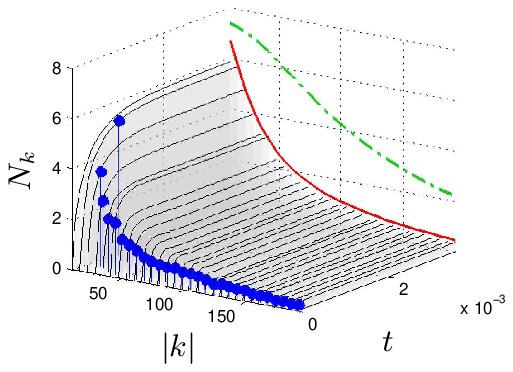}}
		\hspace{5mm}
		\subfigure[$\; \taux_s = 5 \times 10^{-1}$]{\includegraphics[width=0.33\textwidth,clip]{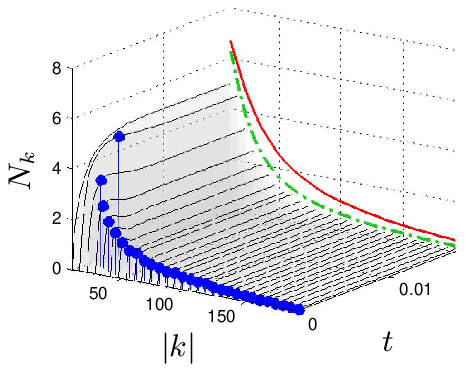}}
		}
		\caption{$\tanh$ expansion: time dependence of Bogoliubov mode populations. Parameters are $C_{NL}(\overline{t} = 0) = 1 \times 10^5$, $N_0 = 10^7$ and $X = 10^3$. The blue points on each curve show where each mode crosses over from phonon to free-particle behaviour (as defined by $\overline{k}_c^2/2 = C_{NL}$). The green dashed curve shows the analytic prediction in the acoustic approximation from (\ref{Nk_tanh}). The red solid curve shows the analytic prediction (including quantum pressure) for a sudden transition from (\ref{Nout_sudden3}).}\label{fig:tanhresults}
		\end{center}
\end{figure*}
%

%% tanh results - large versions for Physical Review A
%\begin{figure*}[!t]
%    \begin{center}
%		\mbox{
%		\subfigure[$\; \taux_s = 1 \times 10^{-1}$]{\includegraphics[width=0.4\textwidth]{Figures/Tanh/modes_tanh_eta0_1e-1.eps}}
%		\hspace{0mm}
%		\subfigure[$\; \taux_s = 5 \times 10^{-1}$]{\includegraphics[width=0.36\textwidth]{Figures/Tanh/modes_tanh_eta0_5e-1.eps}}
%		}
%		\caption{$\tanh$ expansion: time dependence of Bogoliubov mode populations. Parameters are $C_{NL}(\overline{t} = 0) = 1 \times 10^5$, $N_0 = 10^7$ and $X = 10^3$. The blue points on each curve show where each mode crosses over from phonon to free-particle behaviour (as defined by $\overline{k}_c^2/2 = C_{NL}$). The green dashed curve shows the analytic prediction in the acoustic approximation from (\ref{Nk_tanh}). The red solid curve shows the analytic prediction (including quantum pressure) for a sudden transition from (\ref{Nout_sudden3}).}\label{fig:tanhresults}
%		\end{center}
%\end{figure*}
%%

For an expansion corresponding to the de Sitter universe, the scaling function $b(t)$ takes the form (\ref{bdesitter}).
An intuitive picture of the effect of quasiparticle production is demonstrated by Fig.~\ref{fig:desittersnapshot}, which gives the field density at the initial and final times for the case of $t_s = 1 \times 10^{-5}$. The small scale fluctuations given by the initial quasiparticle (Bogoliubov) vacuum are amplified to a larger scale after expansion has occurred.

Figure \ref{fig:desitterresults} shows the results for an expansion of $X = 2000$ and four different rates of expansion $t_s = 1 \times 10^{-5}, 5 \times 10^{-5}, 1 \times 10^{-4}$ and $1 \times 10^{-3}$. In particular, the mode populations are shown as a function of $|k|$ and time. 

For comparison, the sudden transition result for $X = 2000$ from (\ref{Nout_sudden2}) is shown at the final time by the red dashed curve. This gives the upper limit (including the effect of the quantum pressure)
on the permissible particle production in each mode. The green dashed curve shows the analytic prediction at the final time, that is calculated in the acoustic approximation using Bogoliubov theory and Hamiltonian diagonalization as outlined in Sec.~\ref{sect:suddenacoustic}.
Also shown is the time $t_c$ when each mode crosses from phonon to particle-like behaviour due to the expansion of the universe, as determined by $k^2/2 = C_{NL}(t_c)$ -- this is indicated by the blue points on each plot.

\begin{figure}[!t]
\begin{center}
\includegraphics[width=0.85\columnwidth]{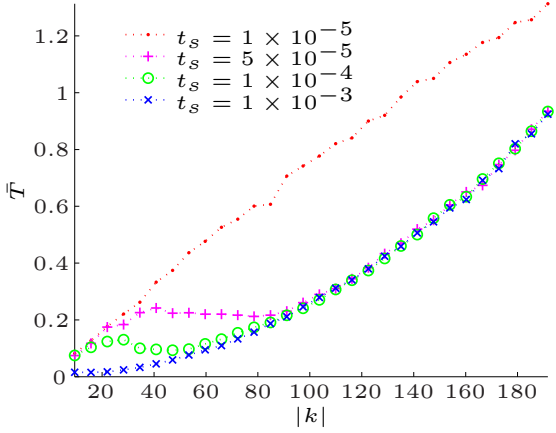}
\caption{Equipartition of energy at the final time for de Sitter expansion with four different expansion rates. Parameters are $C_{NL}(t = 0) = 1 \times 10^5$, $N_0 = 10^7$ and $X = 2 \times 10^3$.}\label{fig:equipartition}
\end{center}
\end{figure}

%+++++++++++++++++++++++++++++++++++++++++++++++++++++++++++++++++
\subsection{$\tanh$ Expansion}
%+++++++++++++++++++++++++++++++++++++++++++++++++++++++++++++++++
In this case the scaling function $b(t)$ is given by the parametric curve in $\taux$ determined by equations (\ref{btaux}) and (\ref{ttaux}). The form of $b(t)$ tends to exhibit a rapid early change followed by a long tail as it approaches its final value $b(t_f)$. Due to the computational expense in running simulations for long times, we took the final simulation time as $\taux_f = \taux_s \tanh^{-1}(0.999)$. The final value of the scaling function $b(t_f)$ then reached  $\approx 99.9\%$ of its target value $1/X$. This minor discrepancy had a negligible effect on the results shown.

Figure \ref{fig:tanhresults} shows the results for an expansion of $X = 1000$ and two different rates of expansion, $\taux_s = 0.1$ and $0.5$. 
Similarly to the de Sitter expansion results, the sudden result for $X = 1000$ is shown by the red solid curve and the phonon to particle-like crossover is indicated by the blue points. Moreover, the green dotted curve shows the particle production for each mode at the final time according to the prediction for the acoustic approximation given by (\ref{Nk_tanh}).

%--------------------------------------------------------
% cyclic results
\begin{figure*}[!htb]
    \begin{center}
		\mbox{
		\subfigure[$\; m = 1$, $X = 2000$, $t_s = 1 \times 10^{-5}$]{\includegraphics[width=0.35\textwidth,clip]{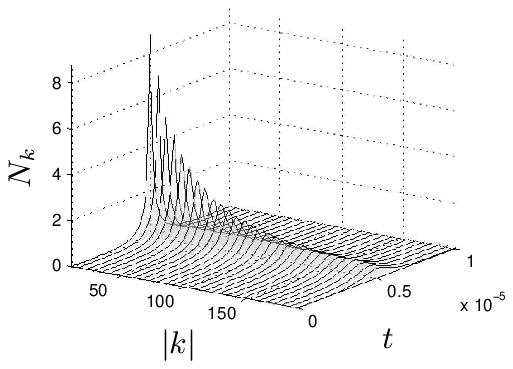}}
		\hspace{5mm}
				\subfigure[$\; m = 1$, $X = 2000$, $t_s = 1 \times 10^{-4}$]{\includegraphics[width=0.35\textwidth,clip]{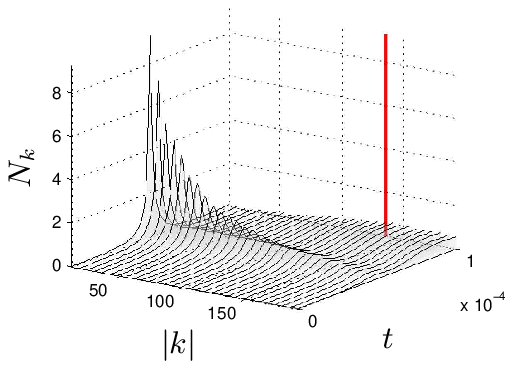}}
		}
		\mbox{
		\subfigure[$\; m = 5$, $X = 100$, $t_s = 1\times 10^{-4}$]{\includegraphics[width=0.35\textwidth,clip]{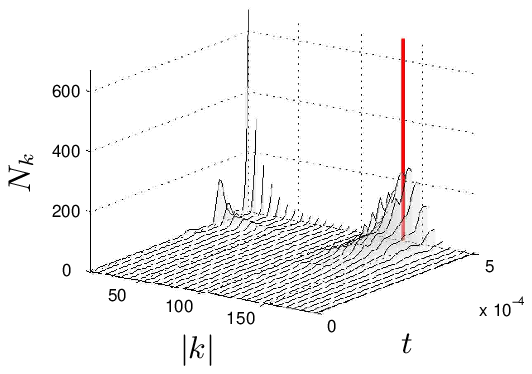}}
		\hspace{5mm}
		\subfigure[$\; m = 10$, $X = 2$, $t_s = 1\times 10^{-4}$]{\includegraphics[width=0.35\textwidth,clip]{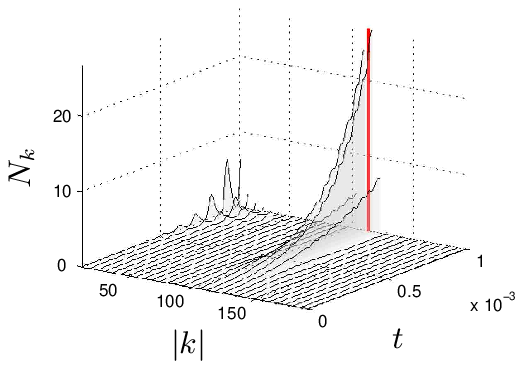}}			
		}
		\caption{Cyclic universe: Time dependence of Bogoliubov mode populations for four different scenarios. Parameters are $C_{NL}(\overline{t} = 0) = 1 \times 10^5$, $N_0 = 10^7$ for all cases. The red solid line shows the position of the peak wave-vector for parametric resonance from the analytic prediction (\ref{cyclicresonance}). In subplots (a) and (b), the spike in population at $t \sim 0.5 t_f$ is an artifact of the time-dependent quasiparticle projection.
		}\label{fig:cyclicresults}
		\end{center}
\end{figure*}
%
%
%% cyclic results - large eps for Physical Review A
%\begin{figure*}[!htb]
%    \begin{center}
%		\mbox{
%		\subfigure[$\; m = 1$, $X = 2000$, $t_s = 1 \times 10^{-5}$]{\includegraphics[width=0.4\textwidth]{Figures/cyclic/cyclic_m1_X2000_C1e5_N1e7_ts_1e-5.eps}}
%		\hspace{0mm}
%				\subfigure[$\; m = 1$, $X = 2000$, $t_s = 1 \times 10^{-4}$]{\includegraphics[width=0.4\textwidth]{Figures/cyclic/cyclic_m1_X2000_C1e5_N1e7_ts_1e-4.eps}}
%		}
%		\mbox{
%		\subfigure[$\; m = 5$, $X = 100$, $t_s = 1\times 10^{-4}$]{\includegraphics[width=0.4\textwidth]{Figures/cyclic/cyclic_m5_X100_C1e5_N1e7_modes.eps}}
%		\hspace{0mm}
%		\subfigure[$\; m = 10$, $X = 2$, $t_s = 1\times 10^{-4}$]{\includegraphics[width=0.4\textwidth]{Figures/cyclic/cyclic_m10_X2_C1e5_N1e7_modes.eps}}			
%		}
%		\caption{Cyclic universe: Time dependence of Bogoliubov mode populations for four different scenarios. Parameters are $C_{NL}(\overline{t} = 0) = 1 \times 10^5$, $N_0 = 10^7$ for all cases. The red solid line shows the position of the peak wave-vector for parametric resonance from the analytic prediction (\ref{cyclicresonance}). In subplots (a) and (b), the spike in population at $t \sim 0.5 t_f$ is an artifact of the time-dependent quasiparticle projection.
%		}\label{fig:cyclicresults}
%		\end{center}
%\end{figure*}
%
%+++++++++++++++++++++++++++++++++++++++++++++++++++++++++++++++++
\subsection{Cyclic universe}
%+++++++++++++++++++++++++++++++++++++++++++++++++++++++++++++++++
In this case, we consider the oscillating scaling function given by (\ref{eqn:scalecyclic}). In particular, we consider two subcases: a single cycle ($m = 1$), which corresponds to a single expansion and contraction for the effective spacetime; and multiple cycles ($m > 1$), which correspond to $m$ expansion and contractions. The peak wave-vector $k_{\textrm{res}}$ for parametric resonance from (\ref{cyclicresonance}) is shown by the position of the red solid line in each case. 

%--------------------------------------------------------
\subsubsection{Single cycle $($$m = 1$$)$}

We further consider two specific sets of parameters: (i) $m = 1$, $X = 2000$ and $t_s = 1 \times 10^{-5}$; and (ii) $m = 1$, $X = 2000$ and $t_s = 1 \times 10^{-4}$. The corresponding simulation results are given by Figs. \ref{fig:cyclicresults}(a) and (b). In both cases there is a transient phase where there is dramatic quasiparticle production in the time-dependent Bogoliubov basis; however, the net quasiparticle production at the end of the cycle ($t = t_f$) is small in both cases, and in particular is very close to zero for the faster cycle in Fig.~\ref{fig:cyclicresults}(a). Note that for $t_s = 1 \times 10^{-5}$ the resonance condition occurs for a wave-vector larger than the cut-off for the projector (\emph{i.e.}, $k_{\textrm{res}} > k_{\textrm{cut-off}}$).

\subsubsection{Multiple cycles $($$m > 1$$)$}

The excitation of the field due to parametric resonance is much greater when there are multiple cycles of the scaling factor, and we can therefore consider a smaller driving amplitude $X$. In particular, we consider two specific sets of parameters for this subcase: (i) $m = 5$,  $X = 100$ and $t_s = 1 \times 10^{-4}$; and (ii) $m = 10$, $X = 2$ and $t_s = 1 \times 10^{-4}$. The corresponding simulation results are given by Figs. \ref{fig:cyclicresults}(c) and (d). Clearly the mode population is peaked near the resonance condition.

%------------------------------------------------------------------------------
\section{Discussion}\label{sect:discussion}

\subsection{Inflationary universe models}

The results for the de Sitter and $\tanh$ cases can be interpretted as follows.
The healing length increases as the expansion proceeds; a mode $k$ that starts as phononic ($k < 1/\xi(0)$) will at a later time $t_c$ cross-over to a particle-like regime ($k \sim 1/\xi(t_c)$). According to the discussion of Sec.~\ref{sect:quantisation} we expect particle production to be dominant before this time, the mode populations becoming fixed after this time. This prediction is bourne out in the results shown in Figs. \ref{fig:desitterresults} and \ref{fig:tanhresults}. 

\begin{figure*}[!htb]
    \begin{center}
		\mbox{
		\subfigure[$\; t = 0.25 t_f$]{\includegraphics[width=0.35\textwidth]{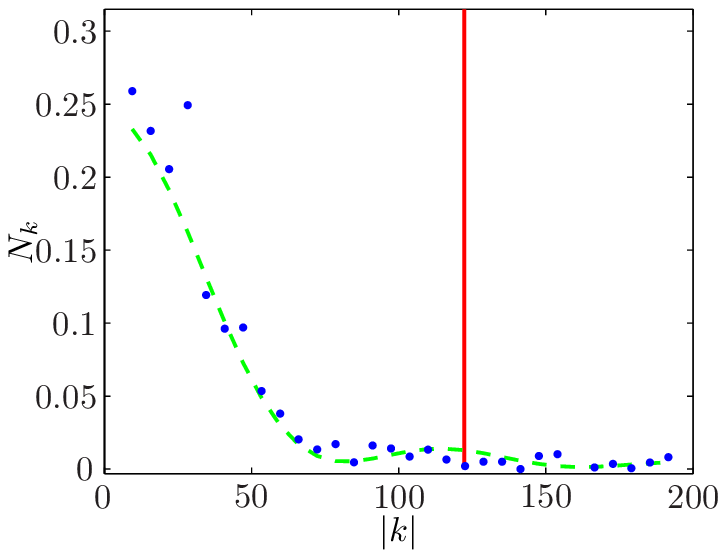}}
		\hspace{5mm}
		\subfigure[$\; t = 0.5 t_f$]{\includegraphics[width=0.35\textwidth]{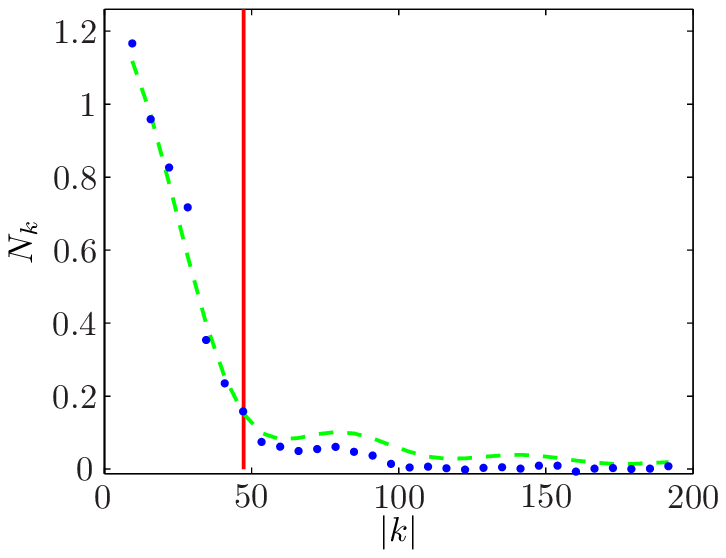}}
		}
		\mbox{
		\subfigure[$\; t = 0.75 t_f$]{\includegraphics[width=0.35\textwidth]{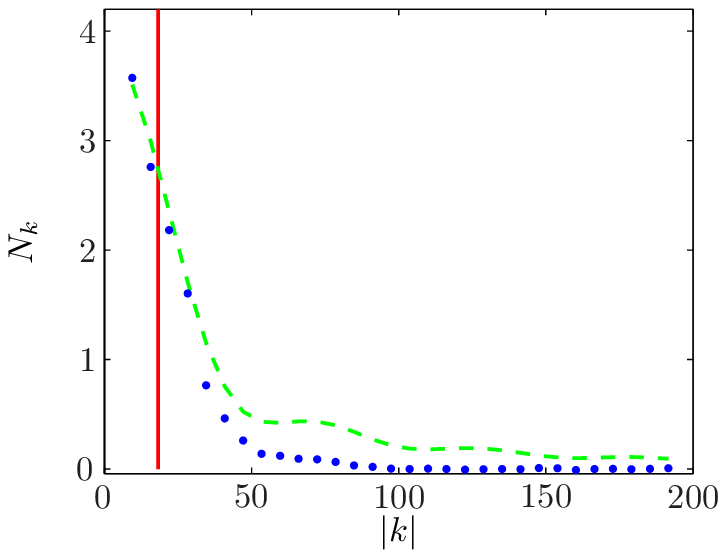}}
		\hspace{5mm}
		\subfigure[$\; t = t_f$]{\includegraphics[width=0.35\textwidth]{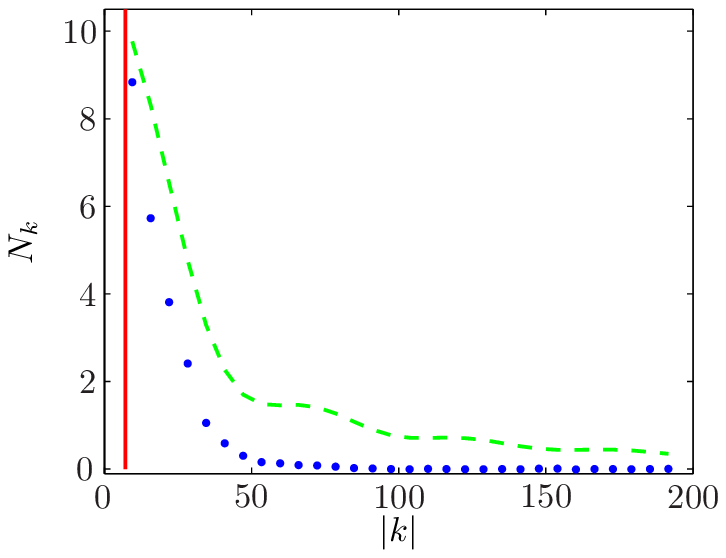}}		}
		\caption{de Sitter expansion: Bogoliubov mode populations at four different times. Parameters are $t_s = 1 \times 10^{-4}$,  $C_{NL}(\overline{t} = 0) = 1 \times 10^5$, $N_0 = 10^7$ and $X = 2 \times 10^3$ at $t_f$. The blue points are the simulation points; the vertical solid red line shows the crossover from phonon to free-particle behaviour (as defined by $\overline{k}_c^2/2 = C_{NL}$); and the green dashed curve shows the analytic prediction in the acoustic approximation based on the Bogoliubov theory and Hamiltonian diagonalization as outlined in Sec.~\ref{sect:suddenacoustic}.}\label{fig:desitterresults2}
		\end{center}
\end{figure*}

For fast expansions, the analytic prediction in the acoustic approximation overestimates the quasiparticle production for modes that cross-over from phonon-like to particle like.  
The role of the cross-over is further illustrated in Fig.~\ref{fig:desitterresults2}. Here we have plotted the quasiparticle numbers $N_k$ at four different times, corresponding to the results from Fig.~\ref{fig:desitterresults}(c) for $t_s = 1 \times 10^{-4}$. For each time, the blue points represent the calculated $N_k$ from the classical field simulations, the green dashed curve is the analytic prediction in the acoustic approximation from Bogoliubov Hamiltonian diagonalization, and the solid red vertical line represents the value of the cross-over $k_c$. As can be seen, for modes that are phononic (left of $k_c$), the quasiparticle production from the classical field simulations closely match the analytic prediction for the acoustic approximation. As the expansion proceeds $k_c$ decreases and $N_k$ for modes $k > k_c$ is suppressed below the analytic prediction. 

Moreover, as is clear from Figs. \ref{fig:desitterresults} and \ref{fig:tanhresults}, the sudden transition prediction given by (\ref{Nout_sudden2}) sets an upper limit on the particle production in each mode. A sudden transition corresponds to $t_s \rightarrow 0$ in the de Sitter expansion case, or $\taux_s \rightarrow 0$ in the $\tanh$ expansion case; in this limit the classical field does \emph{not} evolve, but the Bogoliubov basis that diagonalizes the Hamiltonian changes. The results clearly indicate that the particle production approaches the sudden transition prediction in both scenarios when the expansion rate is largest --- see Figs. \ref{fig:desitterresults}(a) and \ref{fig:tanhresults}(a). The discrepancy for small $|\mathbf{k}|$ in the $\tanh$ expansion case is due to nonlinear interactions, which are neglected in the free field theory of Sec.~\ref{sect:quantisation}. We elaborate on this point below. 

%------------------------------------------------------------------------------
\subsection{Thermal equilibrium and the adiabatic regime}

There are two factors which can lead to a thermal spectrum for the occupation numbers of quasiparticles. 

Firstly, if the expansion is adiabatic, the particle production leads to a thermal spectrum \cite{Birrell1982}. This is the case even when the underlying theory is a free field as in Sec.~\ref{sect:emergentFRW}. 

Secondly, as is the case with our simulations, the CFM is based on the field dynamics for the Hamiltonian (\ref{secondquantH}) which implicitly includes higher-order terms not present in the approximate Bogoliubov Hamiltonian (\ref{eq:Hbog}); therefore the Bogoliubov modes are interacting --- albeit weakly --- and even in the absence of damping the system will eventually approach thermal equilibrium due to ergodicity. 

In the CFM the temperature for a weakly interacting system can be estimated by assuming equipartition of energy.  Following Davis \etal~\cite[see Sec.~VI]{Davis2002}, this gives:
\begin{eqnarray}\label{equipart1}
\langle N_{\mathbf{k}} \rangle_{W} = \frac{k_B T}{E_{\mathbf{k}} - \mu},
\end{eqnarray}
where $E_{\mathbf{k}} = \epsilon_{\mathbf{k}} + \lambda$ is the energy for each Bogoliubov mode in a condensate with eigenvalue $\lambda$, and $\mu$ is the chemical potential. Here $\langle N_{\mathbf{k}} \rangle_{W}$ refers to the quasiparticle population calculated in the Bogoliubov basis.
We can then write
\begin{eqnarray}\label{equipart2}
\epsilon_{\mathbf{k}} = k_B T \left ( \frac{1}{\langle N_{\mathbf{k}} \rangle_{W}} - \frac{1}{\langle N_0 \rangle_{W}} \right ).
\end{eqnarray}
Defining the dimensionless temperature as $\bar{T} = k_B T / \epsilon_L N$, where $\epsilon_L = m L^2 / \hbar^2$, and rearranging (\ref{equipart2}) gives
\begin{eqnarray}\label{temperature_equipart2}
\bar{T} =  \bar{\epsilon}_{\mathbf{k}} \left ( \frac{N}{\langle N_{\mathbf{k}} \rangle_{W}} - \frac{N}{\langle N_0 \rangle_{W}} \right )^{-1}.
\end{eqnarray}
That is, if the system is in thermal equilibrium the function $\bar{T}(|\mathbf{k}|)$ should be constant.

%------------------------------------------------------------------------------%

In Fig.~\ref{fig:equipartition} we plot this function for the de Sitter expansion results with $t_s = 1 \times 10^{-5}, 5 \times 10^{-5}$, $1 \times 10^{-4}$ and $1 \times 10^{-5}$, and at the final time in each case for $X = 2000$. For the fastest expansion with $t_s = 1 \times 10^{-5}$, $\bar{T}$ is approximately linear, indicating the system is not in thermal equilibrium. This is expected since the fastest expansion rate approaches the sudden expansion case, for which the particle production (\ref{Nout_sudden3}) is not thermal. For the slowest expansion with $t_s = 1 \times 10^{-3}$, there is negligible particle production so that the mode populations are fixed at the initial value of half a particle per mode for the classical field. In this case it follows from (\ref{temperature_equipart2}) that $\bar{T} \sim \bar{\epsilon}_{\mathbf{k}}$ as evident in the plot.

In contrast, we note for the two intermediate expansion rates ($t_s = 5 \times 10^{-5}$ and $1 \times 10^{-4}$), $\bar{T}$ is relatively flat for small $|\mathbf{k}|$ which corresponds to the regions in Fig.~\ref{fig:desitterresults} where the particle production is most significant. In these cases, the slower expansions result in an approximately thermal spectrum for the phononic modes, which is consistent for adiabatic expansion in the free-field theory. 
For the larger $|\mathbf{k}|$ modes, no particle production occurs, and the mode populations are frozen at the initial value of half a particle per mode in the classical field. If the field was further evolved at the final nonlinearity $C_{NL}(t = 0)/X$, the system should eventually reach thermal equilibrium via ergodicity. This effect is evident in the $\tanh$ expansion results (Fig.~\ref{fig:tanhresults}) where the nonlinearity $C_{NL}$ asymptotically approaches a non-zero final value. In particular, the nonlinear mode mixing accounts for the discrepancy between the analytic predictions for the free-field theory and the particle production in the low $|\mathbf{k}|$ modes. This effect is more pronounced in Fig.~\ref{fig:tanhresults}(b) where the system evolves for a much longer time. 

\subsection{Cyclic universe model}

The mode spectrum that results from the implementation of a cyclic universe model is markedly different to the case of inflationary expansion (\emph{i.e.}, de Sitter, $\tanh$ or sudden expansions). Specifically, there is a non-zero wave-vector at which the mode population peaks, which is given by the condition for parametric resonance as discussed in Sec.~\ref{sect:cyclicacoustic}. The results for a cyclic universe are given for a single cycle by Figs. \ref{fig:cyclicresults}(a) and (b), and for multiple cycles by Figs. \ref{fig:cyclicresults}(c) and (d). 

For the subcase of a single cycle ($m = 1$) there are two observable effects: 
\begin{itemize}
\item[(i)] A peak in quasiparticle number midway through the cycle, which corresponds to the usual notion of particle production due to expansion as in the inflationary models. This is an artifact of projecting the quasiparticle number into the time-dependent Bogoliubov basis with nonlinearity $U = U_0/X$ (\emph{i.e.}, Hamiltonian diagonalization). 
\item[(ii)] The net quasiparticle number at the end of the cycle is determined by projecting into the Bogoliubov basis with nonlinearity $U(t_f) = U(0)$. That is, the effective spacetime is the same at the start and end of the cycle (providing quasiparticle production does not lead to appreciable depletion of the condensate). Any quasiparticle production is then attributed to parametric excitation, which peaks for the wave-vector corresponding to the resonance condition (\ref{cyclicresonance}). For the cycle with a longer period ($t_s = 1 \times 10^{-4}$), shown in Fig.~\ref{fig:cyclicresults}(b), the resonant wave-vector is within the projected mode-space (\emph{i.e.}, $k_{\textrm{res}} < k_{\textrm{cut-off}}$) and there is non-zero quasiparticle production at the end of the cycle. In contrast, for the cycle of shorter period ($t_s = 1 \times 10^{-4}$), shown in Fig.~\ref{fig:cyclicresults}(a), the resonant condition occurs for modes outside the projected mode-space (\emph{i.e.}, $k_{\textrm{res}} > k_{\textrm{cut-off}}$) and there is negligible quasiparticle production for the system modes. 
\end{itemize}

Quasiparticle production is more dramatic for the case of multiple cycles as we can see from the results in Figs. \ref{fig:cyclicresults}(c) and (d). In particular, referring to the discussion in Sec.~\ref{sect:cyclicacoustic}, for the first case with $m = 5$ and $X = 100$, the parametric resonance leads to a broad peak which can be attributed to the large value of $X$.  By contrast, for the second case with $m = 10$ and $X = 2$ the peak is narrower due to the smaller value of $X$. A thermal component is also evident in both cases for low momenta modes, due to interactions between the quasiparticle modes. This component is expected to grow with longer evolution times as the modes continue to interact (noting that $C_{NL}$ is signficant throughout the evolution).

%%%%%%%%%%%%%%%%%%%%%%%%%%%%%%%%%%%%%%%%%%%%%%%%%%%
%
\section{Conclusions}\label{conclusions}
%
%%%%%%%%%%%%%%%%%%%%%%%%%%%%%%%%%%%%%%%%%%%%%%%%%%%
In summary, we have run classical field simulations for analogue models of inflationary cosmology (de Sitter and $\tanh$ expansions) and also for a cyclic universe model. For the inflationary models the calculation of quasiparticle production $N_k$ shows the following trends:
\begin{itemize}
\item[(i)] $N_k$ is enhanced for faster expansions (small $t_s$) and larger expansions (large $X$). In the limit of a very fast expansion, the results approach the sudden result from (\ref{Nout_sudden3}), which is expected since the field does not evolve for a sufficiently fast change in the nonlinearity. Quasiparticle production is always suppressed below the sudden prediction.
\item[(ii)] $N_k$ is larger for small momenta, since each mode is strongly \emph{coupled} to the effective-spacetime in this case. Alternatively stated, the field equation (\ref{fieldequation_f}) becomes adiabatic for large momenta, so that these modes are not strongly excited by the expansion. Hence the classical field simulations demonstrate the effects of Lorentz violation for the effective spacetime. As expected, the analytic predictions within the acoustic approximation agree favourably with the quasiparticle production for small momenta. 
\end{itemize}

Finally, for the case of a cyclic universe, quasiparticle production is entirely attributed to parametric excitation of the modes as there is no net change to the effective spacetime. In particular, parametric resonance is observed for a mode satisfying $\omega(k) = \Omega/2$ where $\Omega$ is the driving frequency. The parametric resonance peak broadens in momentum space for larger driving amplitude $\Delta$ as expected from the analysis in \cite{GarciaRipoll1999}.

Our calculations (both analytical and numerical) clearly indicate that quasiparticle production should occur in a number of different scenarios, but it is worth commenting on relevant experimental studies that have appeared in the literature, and to then suggest possible extensions to our simple model.  

%+++++++++++++++++++++++++++++++++++++++++++++++++++++++++++++++++
\subsection{Possible experimental implementation}
%+++++++++++++++++++++++++++++++++++++++++++++++++++++++++++++++++
While a specific experiment corresponding to the analogue FRW model we have described has not yet been implemented, there has been significant progress in experiments which might ultimately lead to this goal. In particular, the two main features required by our model are: (i) a homogeneous BEC in a box trap; and (ii) a time-varying scattering length. 

In particular, some progress has been made in experimental implementations of a box trap for BECs, including a square well potential with high barriers on an atom chip  \cite{Hansel2001} and a novel optical trap \cite{Meyrath2005}. It should be noted, however, in both these experiments the resulting hard-wall potential confined the condensate in only one dimension.  On the other hand, the implementation of a time-varying scattering length is easily achievable in a number of different atomic species by making use of a Feshbach resonance \cite{Vogels1997,Inouye1998}. As noted in Sec.~\ref{sect:cfmsuitable}, a promising candidate in this regard is $^{85}\textrm{Rb}$ which allows widely tunable interactions \cite{roberts1998,cornish2000}. However, it should be emphasized that there are additional complications near a Feshbach resonance from either the inelastic processes of three-body recombination (TBR) \cite{Zhang2005}, molecule formation \cite{Donley2002} or the existence of Efimov states \cite{Braaten2003}. It should be possible to avoid these issues in experiments by not tuning the interactions too closely to the Feshbach resonance, or by expanding on a time scale too fast for TBR to have an appreciable effect.

Irrespective of the exact details for the experimental realisation of the FRW analogue model --- \emph{i.e.}, for a homogeneous condensate in a box trap --- we should apply either (or both) of the conditions: (i) that the potential at the edge of the box satisfies $V_{\textrm{box}} \gg U n$, or (ii) that the time-scale of expansion should be very small compared with the trapping frequency so that the condensate remains in the ground state of the trap. 

Finally, it is worth commenting that there is already experimental evidence for excitation of a condensate (\emph{i.e.}, quasiparticle production) due to a time-varying scattering length. In one such experiment by Claussen \etal~\cite{Claussen2002}, Bose condensed $^{85}\textrm{Rb}$ atoms were subjected to an increase in scattering length, followed by a hold time and then a reduction in scattering length. In that work the resulting depletion of the condensate increased with decreasing rise time, except for small hold times ($t_{\textrm{hold}} \leq 15 \mu$s) and small rise times ($t_{\textrm{rise}} \lesssim 20 \mu$s). The time scale over which the scattering length was modified was too small for the condensate shape to adjust dynamically. This experiment would therefore correspond to a contraction and expansion of an effective spacetime, and the dependence of the condensate particle loss on the rise time is consistent with our predictions of quasiparticle production. This effect has been reproduced in numerical simulations by using a generalised Gross-Pitaevskii equation \cite{Duine2003}, although it was necessary to include the effects of TBR there because of the relatively long hold times. 

%+++++++++++++++++++++++++++++++++++++++++++++++++++++++++++++++++
\subsection{Outlook}
%+++++++++++++++++++++++++++++++++++++++++++++++++++++++++++++++++
The FRW analogue model we have considered is based on several simplifying assumptions, the two most significant being homogeneity of the condensate and the two-dimensional box geometry. The preceding discussion therefore motivates several directions in which the formalism could be extended to deal with any realistic experiments that would implement a FRW analogue model of an expanding unverse:

\begin{enumerate}
\item In an experimental implementation of the FRW analogue model, the actual trapping potential may differ from a two-dimensional box trap --- specifically, an implementation will likely require a three-dimensional system with a non-zero potential. The extension of the classical field simulations from two to three dimensions is straightforward since the PGPE (\ref{pgpe}) takes the same form in either case. Moreover, it is straightforward to include a realistic trap into the classical field simulations by specifying the non-zero potential in the PGPE. However, due to the significant increase in size of the mode space for three dimensions, these simulations would necessarily require a considerable computational effort (\emph{i.e.}, running trajectories in parallel on a cluster of workstations).

\item As an alternative to the condensate in a box scenario, we might consider the case of a BEC in a harmonic trap where the trapping frequency and scattering length are simultaneously modified in such a way that the condensate density is approximately constant at the center of the trap. Thus the FRW analogue model we have considered could be  approximately reproduced near the center of the trap.

\item It may be necessary to include the effects of TBR to accurately describe the dynamics close to a Feshbach resonance. The inclusion of TBR into the TWA has been previously described by Norrie \etal~\cite{Norrie2006b}.

\item Finally, the presence of a thermal cloud will certainly affect the results of any experiment, possibly obscuring the signal of quasiparticle production due to expansion. Finite temperature effects may be included by using the classical field method whereby the phase space is separated into a \emph{coherent} region (highly occupied modes) and an \emph{incoherent} region (weakly occupied modes). This has been formalised in terms of either the finite temperature GPE \cite{Davis2002} or the stochastic GPE \cite{Gardiner2002}. 

\end{enumerate}

Whether or not any of the above modifications are incorporated into the model, however, one should remain careful that the analogy to a FRW-type universe is preserved in some regime of interest.

%%%%%%%%%%%%%%%%%%%%%%%%%%%%%%%%%%%%%%%%%%%%%%%%%%%
%%%%%%%%%%%%%%%%%%%%%%%%%%%%%%%%%%%%%%%%%%%%%%%%%%%
\begin{acknowledgments}
  This work was supported by Victoria University of Wellington through
  two PhD completion scholarships (PJ and SW).  PJ was also supported by
  the New Zealand Tertiary Education Commission scholarship TAD-1054.
  SW was also supported by a Victoria University Small Research Grant.
  CG and MV were supported by two Marsden Fund grants administered by
  the Royal Society of New Zealand. PJ wishes to thank Ashton Bradley, Craig Savage,
  Pierfrancesco Buonsante and Blair Blakie for useful discussions.
\end{acknowledgments}
%%%%%%%%%%%%%%%%%%%%%%%%%%%%%%%%%%%%%%%%%%%%%%%%%%%%
%%%%%%%%%%%%%%%%%%%%%%%%%%%%%%%%%%%%%%%%%%%%%%%%%%%
%\bibliographystyle{prsty}
%\bibliography{References/analogmodels_expanding,References/analogmodels_common,References/analogmodels_laval,References/references_general,References/Silke}
%\input{FRWCFM_V1c.bbl}
\bibliography{FRWCFM_V1c}
%%%%%%%%%%%%%%%%%%%%%%%%%%%%%%%%%%%%%%%%%%%%%%%%%%%
%%%%%%%%%%%%%%%%%%%%%%%%%%%%%%%%%%%%%%%%%%%%%%%%%%%
\end{document}